\documentclass{article}
\usepackage{graphicx} 
\usepackage{mathrsfs}
\usepackage{amsfonts}
\usepackage{amsmath}
\usepackage{amsthm}
\usepackage{amssymb}
\usepackage{ragged2e}
\usepackage{fancyhdr}
\usepackage{comment}
\usepackage{float}
\usepackage{url}
\usepackage{bm}
\usepackage{optidef}
\usepackage{subcaption}
\usepackage[a4paper, portrait, margin=1in]{geometry}
\usepackage[style=apa]{biblatex}
\usepackage{multirow}
\usepackage[title, toc, page, titletoc]{appendix}
\usepackage{lastpage}
\usepackage[colorlinks=true,linkcolor=blue,citecolor=blue,urlcolor=blue]{hyperref}
\usepackage{listings}
\usepackage{color}
\usepackage{booktabs}
\usepackage{parskip}
\usepackage{setspace}
\usepackage{fvextra}

\hypersetup{pdfborder=0 0 0}

\newcommand\codify[1]{\textcolor{blue}{\texttt{#1}}}

\definecolor{dkgreen}{rgb}{0,0.6,0}
\definecolor{gray}{rgb}{0.5,0.5,0.5}
\definecolor{mauve}{rgb}{0.58,0,0.82}

\providecommand{\keywords}[1]
{
  \small	
  \textbf{Keywords:} #1
}

\makeatletter
\renewcommand\@date{{%
  \vspace{-2\baselineskip}%

  {\large\centering
  \begin{tabular}{@{}c@{}}
    Justin Philip Tuazon\textsuperscript{1} \quad \quad Joemari Olea\textsuperscript{2} \quad \quad Richelle Ann Juayong\textsuperscript{1}
  \end{tabular}}

  \smallskip

  {\normalsize\textsuperscript{1}Department of Computer Science, University of the Philippines Diliman\linebreak
  \textsuperscript{2}Department of Educational Psychology, University of Texas at Austin}

  \smallskip

  {\large September 2026}
}}
\makeatother

\makeatletter
\def\@maketitle{%
  \newpage
  \null
  \vskip 2em%
  \begin{center}%
  \let \footnote \thanks
    {\LARGE\bfseries\@title \par}%
    \vskip 1.5em%
    {\large
      \lineskip .5em%
      \begin{tabular}[t]{c}%
        \@author
      \end{tabular}\par}%
    \vskip 1em%
    {\large \@date}%
  \end{center}%
  \par
  \vskip 1.5em}
\makeatother

\fancypagestyle{firststyle}
{
   \fancyhf{}
   \fancyfoot[R]{\small Page \thepage\ of \pageref{LastPage}}
   \fancyfoot[L]{\small Preprint \\ The latest version of FactorFlow, V3.7.2, is hosted here: \url{https://factorflow-efa.streamlit.app/}.}
   
}

\title{\Large FactorFlow: A Visual Analytics Workspace with Large Language Model--Assisted Interpretation for Factor Analysis}
\author{}

\begin{document}
\justifying
\maketitle
\thispagestyle{firststyle}

\begin{abstract}
    \noindent In exploratory factor analysis (EFA), one aims to describe latent variables by constructing a factor model based on the relationships among manifest variables. For a model to be useful, it is not enough that it is grounded on data; it must also be meaningful. Hence, in practice, one attempts to interpret different factor models to identify a meaningful, coherent, and theoretically defensible latent structure. Doing so, however, is not straightforward, as it is subjective and requires tracking extensive information. Thus, we introduce FactorFlow, a system designed to help researchers perform EFA more effectively. With an interactive dashboard that supports comprehensively visualizing up to two models simultaneously and large language model integration that enables the generation of automated model interpretations written in natural language, FactorFlow substantially aids the crucial step of model interpretation, all the while supporting the end-to-end workflow. Indeed, our usability survey evidences the effectiveness of FactorFlow.
\end{abstract}


\setcounter{tocdepth}{3}
{\setstretch{1}
\tableofcontents
}

\section{Introduction}
In many cases, researchers and practitioners usually study variables that are \textit{directly} observable or measurable. Examples of such kind of variables include the height of a person in meters, the speed of an animal in kilometers per hour, and the response of a customer to a Likert-type item (e.g., ``Strongly Agree"). As mentioned in \textcite{Mulaik2010}, such variables are called \textit{manifest variables}. However, there are also cases where the variables of interest are \textit{not} directly observable or measurable. For instance, how would you measure the anxiety of a student when it comes to taking exams? How about the underlying ``value consciousness" of a consumer for your business? In contrast, such variables are referred to as \textit{latent variables}, also as described in \textcite{Mulaik2010}. As opposed to manifest variables, latent variables are \textit{abstract} concepts. While we can use a ruler to quickly measure length, we do not have a tool with which we can directly measure ``happiness", for example. Clearly, we need special methods to deal with latent variables.

On this note, there are several approaches that we can take to identify and measure latent variables or phenomena. One such method is \textit{factor analysis}. Usually, factor analysis is used when the manifest variables are both numerical and continuous, but it can also be applied to other cases. Notably, it is often applicable to ordinal data (e.g., Likert-type data), as described in \textcite{ordinal2020}. One can also use specialized correlation matrices or procedures to apply factor analysis to ordinal data \parencite{polychoric}. Given these capabilities, factor analysis has become a widely used method in numerous fields. Indeed, it has become one of the most popular methods in psychometrics, the field where the method originated \parencite{Mulaik2010}. However, beyond psychometrics, factor analysis has also been used in several other fields. For example, \textcite{yilmaz2024} leveraged factor analysis to identify the latent variables that constitute sustainable manure utilization in farming. Meanwhile, \textcite{ledesma2021} provided a review of the application of factor analysis in transportation research (e.g., to study driver behavior) and established guidelines for using such method in the field.

Now, in factor analysis, the latent variables are also called as \textit{factors} and we ultimately build a \textit{factor model}, which relates latent variables to manifest variables. Essentially, factor analysis aims to identify and characterize the factors (i.e., the underlying latent structure) based on the observed relationships (i.e., correlations) among the manifest variables, as discussed in \textcite{Mulaik2010}, and the number of factors is usually much smaller than the number of manifest variables. Moreover, using factor analysis, one can ``measure" the latent variables through estimating \textit{factor scores} based on the factor model \parencite{gorsuch1983}. Here, we focus on \textit{exploratory factor analysis} (EFA), which is primarily a way to come up with hypotheses about the latent structure (i.e., constructing the hypothesized factor model). As its name implies, EFA involves ``exploring" the data and identifying the factors. This is in contrast to \textit{confirmatory factor analysis} (CFA), which is used to \textit{test} the generated hypotheses and establish (i.e., ``confirm") the model and factors, as discussed in \textcite{gorsuch1983}.

Additionally, as an exploratory method, EFA by its nature is an \textit{iterative} procedure, where the researcher fits, rotates, evaluates, and interprets several models until they find and decide on the ``best" one. Furthermore, the success of EFA largely depends on a ``WOW" criterion, where the factor model is ``good" if it is interpretable and ``makes sense" \parencite{stat147Ref}. On the same note, as \textcite{semanticsFA} stated, the one that is more interpretable between two statistically valid factor models is preferred because it provides greater substantive utility and a more meaningful insight into the underlying structure of the data. The goal, ultimately, is to obtain a model that is not only supported by observed data, but is also substantially meaningful. Especially because of the abstract quality of latent variables, claims about factors would be difficult to support if the nature (or description) of the factors themselves are not even understood well enough or well-defined.

Given such, the interpretation of a factor model is arguably the most crucial part of EFA. Interpretation, however, is not a trivial task. First, interpretability cannot be easily quantified and the process of interpretation itself can be subjective. It often relies on the domain, theoretical background, and the subjective judgment of the researcher. Second, interpreting the factor model involves keeping track of multiple pieces of information and looking at them from different perspectives. For instance, if the model has $7$ factors and $50$ manifest variables, one would need to look at $7\times50=350$ values (i.e., loadings) simultaneously. Third, rarely is it the case that the researcher has to interpret only one estimated model. In most cases, if not all, the researcher has to evaluate multiple models (and rotations), compare and contrast them, and eventually select the most meaningful one out of the pool of candidate models.

In connection with this, \textit{visualizations} have become common tools when performing EFA, as they support the critical step of interpreting the (candidate) factor model in the workflow, among other things (e.g., model diagnostics). For example, using a \textit{heatmap} to visualize the model (i.e., the loading matrix) is an effective way of understanding the underlying latent structure. Beyond heatmaps, specialized visualizations, including the classic \textit{scree plot} from \textcite{scree} and the more recent \textit{interpretability plot} from \textcite{tuazon2026}, for EFA have emerged. With the wide usage of visualizations and the exploratory perspective in EFA, a \textit{dashboard} would certainly be a helpful tool when performing the analysis. Indeed, \textcite{favis} demonstrated that bundling various visualizations and providing interactive features for EFA can help the researcher perform the analysis more efficiently and effectively.

Hence, this paper introduces \textit{FactorFlow}, a visual analytics workspace with large language model-assisted interpretation for factor analysis. FactorFlow is an interactive system that provides users with various dynamic visualizations and feature-rich components, including a comprehensive dashboard for interpreting models, for performing EFA end-to-end (i.e., from model estimation and diagnostics to interpretation). This application builds and improves on previous systems or visualizations in three main ways, which are briefly discussed in the next paragraph.

First, FactorFlow covers the end-to-end workflow for EFA, not just the last steps (e.g., interpretation). This means that users can import their dataset, perform initial exploratory data analysis, fit and rotate models, examine model diagnostics, interpret models, and export results and findings all in the same system. Second, our new system contains a dashboard that has a comprehensive set of figures and visualizations, ranging from classical plots to modern visualizations. Most importantly, the dashboard supports the display of \textit{two} factor models at the same time. As mentioned earlier, EFA involves assessing a pool of candidate models to select the ``best" one and as such, the simultaneous side-by-side display of two candidate models can certainly aid the researcher compare models more easily. Third, FactorFlow leverages the power of large language models (LLM) to assist the user with interpreting the factor model. In other words, in addition to the visualizations that already aid interpretation, FactorFlow is able to generate automated interpretations, written in natural language, to directly help the researcher 1) characterize the latent structure and 2) assess how meaningful the model is.

Based on the usability survey conducted in this study, FactorFlow is generally a usable and effective tool. All testers gave positive ratings for the usability of FactorFlow and indicated that they would recommend the tool to others who need to perform EFA. Overall, the testers generally found the tool to be easy-to-use, comprehensive, and effective. Interestingly, one finding from the survey is that FactorFlow can also be used as a \textit{pedagogical} or \textit{teaching} tool for EFA (e.g., for demonstrating the workflow).

The remaining parts of the paper are structured as follows. Section \ref{sec:efa} provides a brief review of the mathematical background and workflow for EFA, as well as a discussion on the use of visualizations in EFA. Meanwhile, Section \ref{sec:ff} formally introduces FactorFlow, including its design and usage. Then, Section \ref{sec:eval} discusses the results and findings the system evaluation (e.g., user assessment). Finally, Section \ref{sec:conc} ends the paper with a summary of the contribution of this paper and directions for future research.

\section{Exploratory Factor Analysis}\label{sec:efa}
\subsection{Background}
At this point, we will briefly go over some foundational concepts in EFA. As described in \textcite{stat147Ref}, the common factor model \begin{math}\mathscr{F}\end{math}, which we estimate in factor analysis, is given by
\begin{displaymath}
\bm{\underline{X}}_{M\times1}=\bm{\underline{\mu}}_{M\times1}+\bm{\underline{L}}_{M\times T}\bm{\underline{F}}_{T\times 1}+\bm{\underline{\varepsilon}}_{M\times1},
\end{displaymath}
where $M$ is the number of manifest variables, $T\left(\ll M\right)$ is the number of factors,
\begin{displaymath}
\bm{\underline{X}}=\begin{pmatrix}
  X_{1} \\
  X_{2} \\
  \vdots \\
  X_{M}
\end{pmatrix}, ~
\bm{\underline{\mu}}=\begin{pmatrix}
  \mu_{1} \\
  \mu_{2} \\
  \vdots \\
  \mu_{M}
\end{pmatrix}, ~
\bm{\underline{L}}=\begin{pmatrix}
  l_{1,1} & \dots & l_{1,T} \\
  \vdots & \ddots & \vdots \\
  l_{M,1} & \dots & l_{M,T}
\end{pmatrix}, ~
\bm{\underline{F}}=\begin{pmatrix}
  F_{1} \\
  F_{2} \\
  \vdots \\
  F_{M}
\end{pmatrix}, ~\text{and}~
\bm{\underline{\varepsilon}}=\begin{pmatrix}
  \varepsilon_{1} \\
  \varepsilon_{2} \\
  \vdots \\
  \varepsilon_{M}
\end{pmatrix}
\end{displaymath}

Here, \begin{math}X_{1},\dots,X_{M}\end{math} are the manifest variables while \begin{math}F_{1},F_{2},\dots,F_{T}\end{math} are the factors. Moreover, \begin{math}\bm{\underline{L}}\end{math} is called the \textit{loading matrix}, with \begin{math}l_{i,j}\end{math} as the loading of \begin{math}X_{i}\end{math} on \begin{math}F_{j}\end{math}. Note that the loading is the covariance (correlation, if the manifest variable was standardized) of the manifest variable with the factor. In line with this, the loading (or factor loading) is a measure of how much the factor generalizes to the manifest variable (in other words, how much the factor ``explains" the manifest variable), as mentioned in \textcite{gorsuch1983}. Finally, \begin{math}\bm{\underline{\mu}}\end{math} is the mean vector of \begin{math}\bm{\underline{X}}\end{math} and \begin{math}\varepsilon_{1},\dots,\varepsilon_{M}\end{math} are the error terms. There are additional assumptions imposed on the variables, which also depend on the class of models considered (e.g., orthogonal factor models). Moreover, there are various estimation methods for fitting the factor model (up to rotation) based on data, such as MinRes from \textcite{Harman1966}. We will not discuss these anymore but details of such can be found in \textcite{Mulaik2010} and \textcite{gorsuch1983}.

Now, as mentioned earlier, the primary goal of EFA is to determine the latent structure. As such, one needs to \textit{interpret} (i.e., assign a meaning to) the factors, which is not a trivial task for reasons explained in the previous section. Operationally, factors are typically interpreted by examining the loading matrix \begin{math}\bm{\underline{L}}\end{math}. When the absolute value of the loading \begin{math}l_{i,j}\end{math} is large, the manifest variable \begin{math}X_{i}\end{math} is considered to be strongly associated with factor \begin{math}F_{j}\end{math}. Thus, the substantive meaning of \begin{math}X_{i}\end{math} can be used to inform the interpretation of \begin{math}F_{j}\end{math}. For example, suppose that all manifest variables come from Likert-type items in a questionnaire (an example questionnaire is shown in Table \ref{tab:example-likert}). To interpret a factor, one looks at the collection of questions that correspond to high-loading manifest variables for that factor. If the questions all reflect anxiety-related content, then the factor may be interpreted as an ``anxiety score" or ``anxiety component". In contrast, if the high-loading items reflect several unrelated constructs, then the factor may lack a coherent substantive interpretation.

\begin{table}[ht]

\centering
\caption{Example Questionnaire with Likert-type Items}
\label{tab:example-likert}
\begin{tabular}{p{0.35\textwidth} c c c c c}
\hline
\textbf{Statement} &
\shortstack{\\ \textbf{1}\\\footnotesize Strongly Disagree} &
\shortstack{\\ \textbf{2}\\\footnotesize Disagree} &
\shortstack{\\ \textbf{3}\\\footnotesize Neutral} &
\shortstack{\\ \textbf{4}\\\footnotesize Agree} &
\shortstack{\\ \textbf{5}\\\footnotesize Strongly Agree} \\
\hline

I feel energized when I start my day. & $\square$ & $\square$ & $\square$ & $\square$ & $\square$ \\
I find it easy to concentrate on tasks. & $\square$ & $\square$ & $\square$ & $\square$ & $\square$ \\
\hline
\end{tabular}

\end{table}

On a related note, note that the factor model \begin{math}\mathscr{F}\end{math} is not \textit{identifiable}, as it is rotationally indeterminate. This means that rotating the loading matrix (i.e., replacing \begin{math}\bm{\underline{L}}\end{math} with \begin{math}\bm{\underline{L}}^{*}=\bm{\underline{L}}\bm{\underline{R}}\end{math}, where \begin{math}\bm{\underline{R}}\end{math} is a rotation matrix) yields an equally valid model with the same model-implied covariance matrix (i.e., an observationally equivalent model). In simpler words, multiple (specifically, infinitely many) estimated models can be validly derived from the same observed dataset. This adds another challenge to the identification and interpretation of the factors, as the researcher needs to generate an orientation of the loading matrix that ``makes sense". In practice, various rotation methods, such as the varimax rotation from \textcite{Kaiser1958}, are applied to the loading matrix to search for the ``best" model.

Moving from theory to practice, what does the EFA workflow look like? Assuming that the collected dataset has been cleaned and is ready for analysis, EFA begins with the estimation of the correlation matrix of the manifest variables, which also involves deciding which manifest variables to include (at the beginning, usually the full set is used but some manifest variables may be dropped after evaluating diagnostics), as mentioned in \textcite{Mulaik2010}. In fact, the common factor model is fitted to the sample correlation matrix, not the raw data per se. This means that even without the raw data, one can fit the model as long as the correlation matrix is available. In general, Pearson correlations are used to compute the matrix. In some cases, however, other correlation types may be more appropriate. For instance, if the observed data are ordinal, the polychoric correlation matrix is more suitable compared to the Pearson correlation matrix \parencite{polychoric}.

Then, one specifies the number of factors. There are several heuristics for identifying the number of factors, some of which are discussed in the next subsection. In any case, after specifying the number of factors, the common factor model (i.e., the initial loading matrix) is estimated through one of many methods, including, but not limited to, MinRes as described in \textcite{Harman1966} and maximum likelihood estimation as described in \textcite{Mulaik2010}. Although, as mentioned, the common factor model \begin{math}\mathscr{F}\end{math} is rotationally indeterminate. In other words, the initial estimated loading matrix can be rotated to produce a new but equally valid loading matrix. Thus, the next and closely related step is to rotate\footnote{One can ``skip" the step of rotating the loading matrix, as the unrotated loading matrix can also be examined. However, this is technically equivalent to rotating with the identity matrix.} the loading matrix (i.e., apply a rotation method such as varimax from \textcite{Kaiser1958}, promax rotation from \textcite{promax}, and so on) to orient it to a (hopefully) more meaningful configuration.

Following that, one can examine model diagnostics. For example, the communalities of the manifest variables can provide insights on how well the factor model fits the data (e.g., small communalities suggest poor fit), as they reflect the portions of the variances of the manifest variable that can be explained by the underlying factors \parencite{gorsuch1983}. In other words, communalities can be used to assess whether some manifest variables (e.g., those with low communalities) need to be removed from the analysis or not. Furthermore, one can examine sampling adequacies by looking at the KMO and MSA indices \parencite{Mulaik2010}. Evaluating the factor correlation matrix may also be beneficial, not just for the later step of interpretation, but also for deciding between orthogonal and oblique models. Although, in some cases, one can skip (parts of) this step, especially if they are concerned only with extracting meaningful factors. Nevertheless, note that most model diagnostics, including communalities and measures of sampling adequacies, can actually be computed and assessed even before rotating the loading matrix. For instance, communalities and measures of sampling adequacies are invariant to factor rotations.

\begin{figure}[H]
    \centering
    \includegraphics[width=1\textwidth]{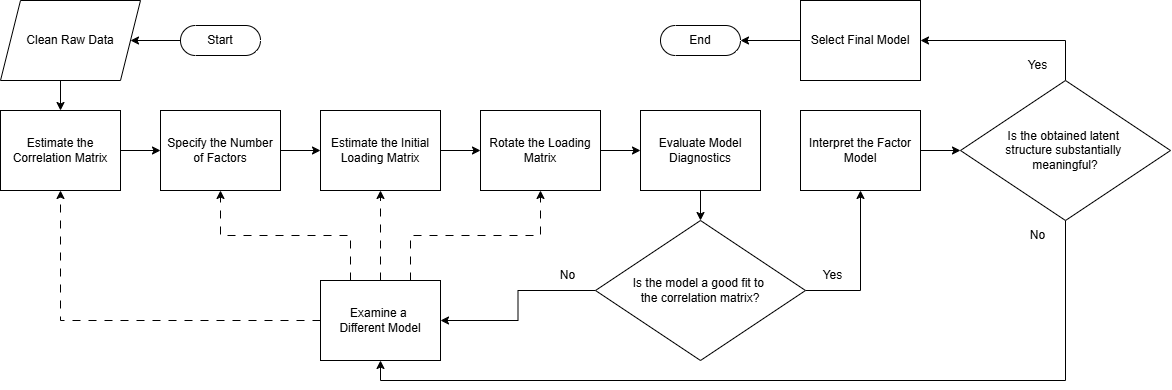}
    \caption{General EFA Workflow}
    \label{fig:efa-workflow}
\end{figure}

Finally, the last step is to interpret the model, primarily by examining the (rotated) loading matrix (as explained previously). This is arguably the most critical step, as the ultimate goal of EFA is to obtain a meaningful (i.e., interpretable, coherent, and theoretically defensible) latent structure, as well as the most tedious one, as it involves the careful manual review of the researcher. In addition, recall that EFA is an exploratory and iterative process. As such, it is often the case that one has to go back to and repeat one or more steps until they appropriately decide on the ``final" model (i.e., the most meaningful candidate model). Usually, one applies different rotation methods, attempts to interpret each resulting candidate model, and compares different models. Given such, the general workflow for EFA is summarized in Figure \ref{fig:efa-workflow}.

\subsection{Visualizations}
Now, as repeatedly mentioned, making (or attempting to make) interpretations based on the loading matrix can be difficult, as it requires identifying patterns from a large number of loadings and looking at the loading matrix from different ``perspectives" (e.g., in terms of magnitude, in terms of trends). For example, if there are $56$ manifest variables and $7$ factors, the number of values (loadings) one has to look at simultaneously is $56\times 7=392$. This becomes even more complicated when the researcher needs to compare multiple candidate models (which is almost always the case). As such, \textit{visualizations} are critical when performing EFA and several types are commonly used.

\begin{figure}[H]
  \centering
  \begin{subfigure}[c]{.5\textwidth}
      \centering
      \includegraphics[width=1\textwidth,height=5cm]{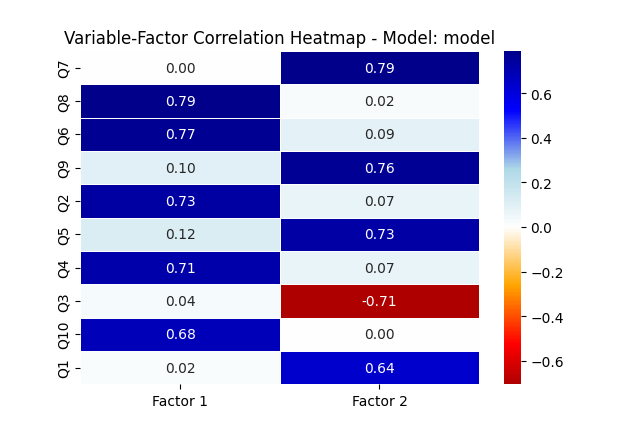}
      \caption{Heatmap of Loadings}
  \end{subfigure}%
  \begin{subfigure}[c]{.5\textwidth}
      \centering
      \includegraphics[width=1\textwidth,height=5cm]{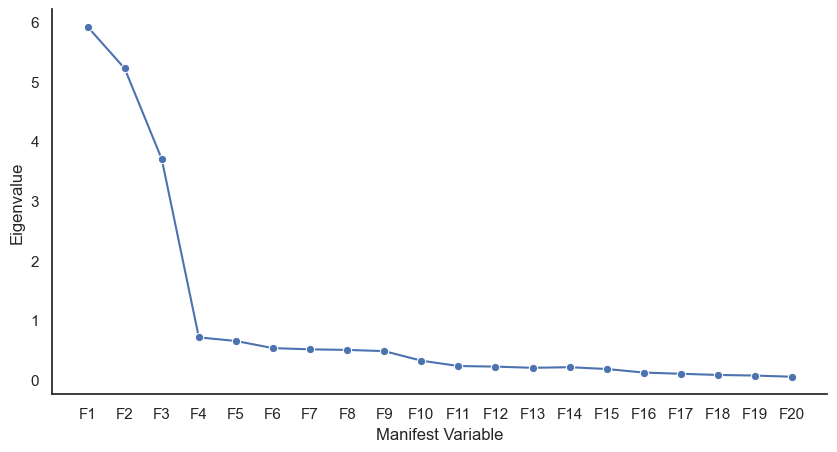}
      \caption{Scree Plot\protect}
  \end{subfigure}
  \begin{subfigure}[c]{.5\textwidth}
      \centering
      \includegraphics[width=1\textwidth,height=5cm]{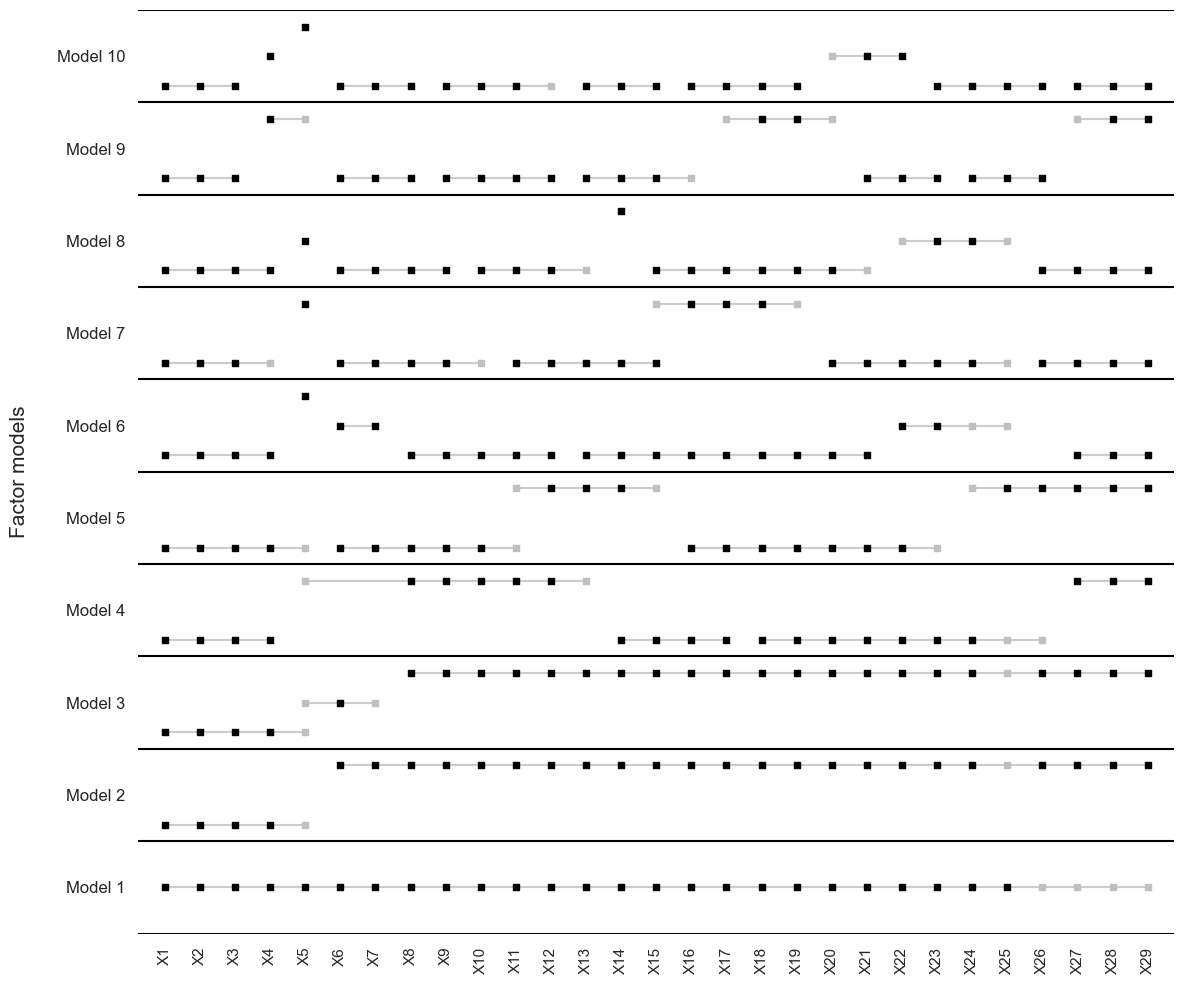}
      \caption{Factor Model Plot\protect}
  \end{subfigure}%
  \begin{subfigure}[c]{.5\textwidth}
      \centering
      \includegraphics[width=1\textwidth,height=5cm]{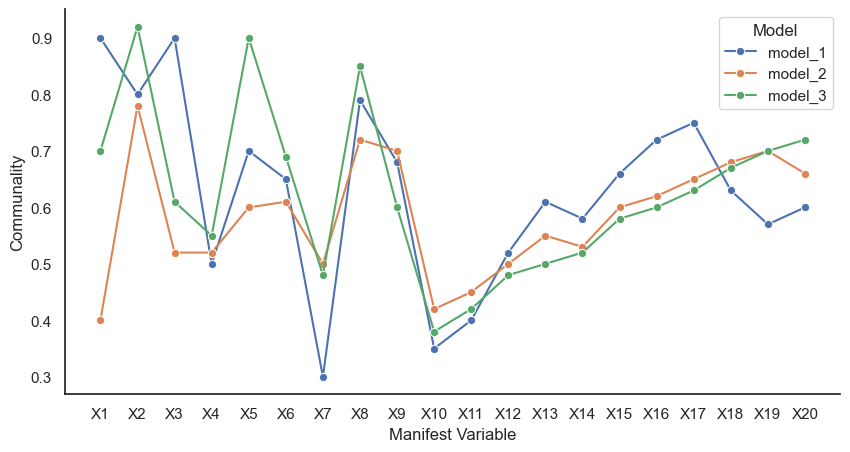}
      \caption{Communality Plot\protect}
  \end{subfigure}
  \caption{Common Visualizations in Factor Analysis}
  \label{fig:example-visualizations-efa}
\end{figure}

Perhaps the most important and common one is the \textit{heatmap}, an example of which is shown in Figure \ref{fig:example-visualizations-efa}(a). The figure shows a heatmap of the loading matrix but heatmaps can also be constructed for other ``parts" of the factor model. For instance, it is common to construct heatmaps for the sample correlation matrix, the communalities, and the sampling adequacies. In the case of heatmaps for loadings, one can ``dichotomize" the loadings such that a loading is colored if its absolute value meets or exceeds a threshold, and is uncolored otherwise. In any case, the heatmap makes it obvious which values are large and which are not. In the case of the loading matrix, this makes it easier to identify which manifest variables should be considered when interpreting the factor.

Another type of visualization commonly used in EFA is the \textit{scree plot} proposed by \textcite{scree}, which provides a heuristic for identifying the ideal number of factors in the model. An example of a scree plot is shown in Figure \ref{fig:example-visualizations-efa}(b). This plot shows the eigenvalues of the (sample) correlation matrix, ordered from largest to smallest. Using the scree plot, one can use several heuristics for deciding on the optimal number of factors. For instance, one can look at the number of eigenvalues that exceed a certain threshold, as in the case of the eigenvalues-greater-than-one rule from \textcite{kaiserRule1,kaiserRule2}, or look at the ``elbow" of the plot based on the criterion from \textcite{scree}, both of which are described in \textcite{Mulaik2010}.

Furthermore, \textcite{vizefa} illustrated the \textit{factor model plot} and the \textit{communality plot}. In the factor model plot, the $x$-axis tick labels refer to the manifest variables and the $y$-axis tick labels refer to the different factor models. In a given row or factor model, the manifest variable is marked with a square if its loading is large enough and squares are connected with segments if they are explained by the same factor. Thus, the factor model plot can make it easy to compare different factor models (or different rotations). In Figure \ref{fig:example-visualizations-efa}(c), variables $X6$, $X7$, $X8$, $X9$, $X10$, and $X11$ are explained by the same factor in factor model $5$. Meanwhile, the communality plot, an example of which is Figure \ref{fig:example-visualizations-efa}(d), shows which manifest variables have low or high communalities, where different lines (distinguished perhaps by color or line style) correspond to different factor models. The communality of a manifest variable is the amount or portion of the variance of the manifest variable that is explained by the factor model. Ideally, a manifest variable should have a high communality.

\begin{figure}[H]
    \centering
    \includegraphics[width=1\textwidth]{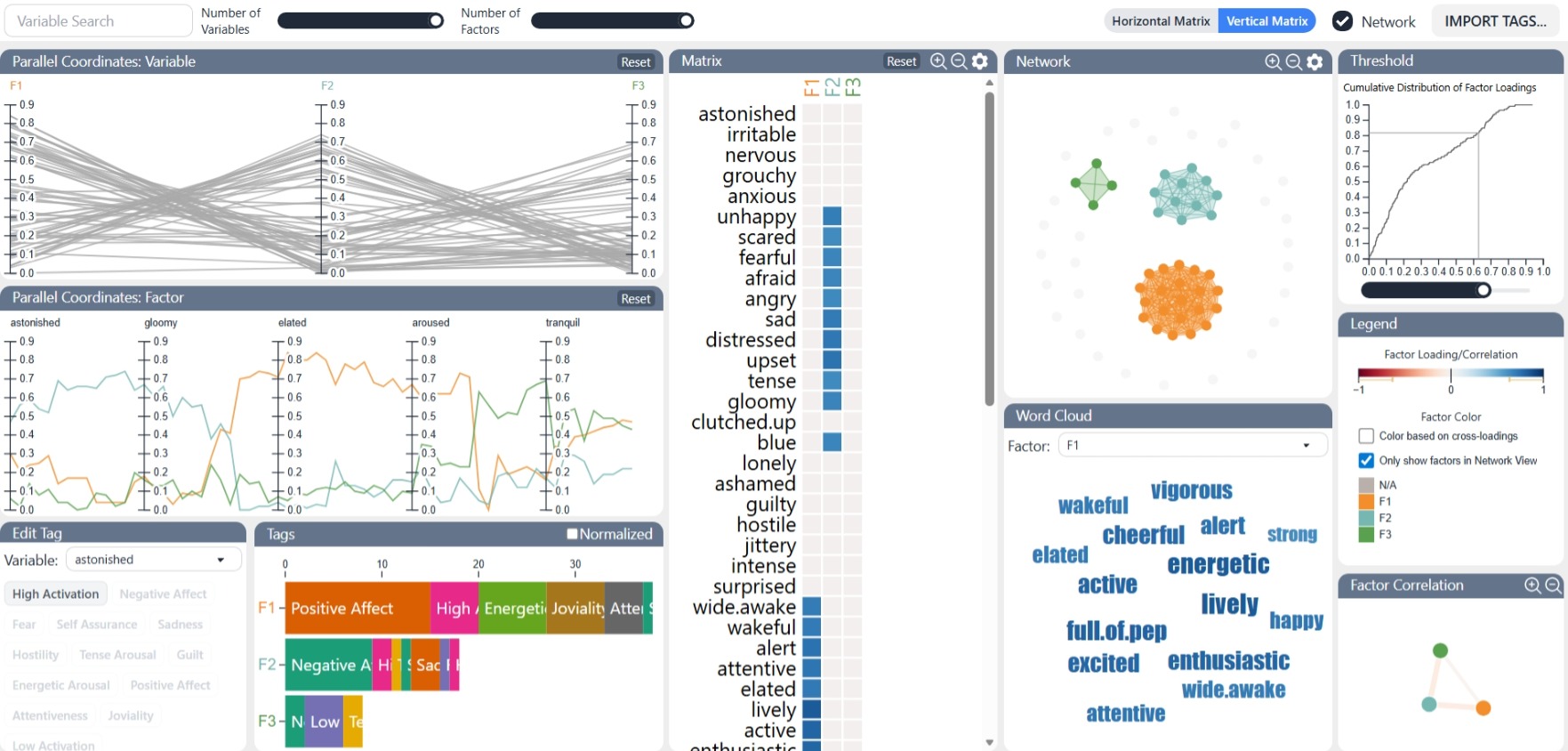}
    \caption{FAVis by \textcite{favis}}
    \label{fig:example-favis}
\end{figure}

Then, given the exploratory nature of EFA and the numerous visualizations employed, it is obvious that a \textit{dashboard} can be useful for performing EFA. In fact, \textcite{favis} developed \textit{FAVis}, a visual analytics dashboard for exploratory factor analysis in psychological research. A screenshot of FAVis is shown in Figure \ref{fig:example-favis}.

\textcite{favis} included \textit{parallel coordinates plots} for the loadings, one for comparing loadings across manifest variables and another for comparing loadings across factors. Using such plots, it is easy to spot manifest variables (or factors) with similar loadings (i.e., those that have parallel or coincident lines are similar), even at a relative scale (e.g., when lines are not necessarily coincident but are parallel). They also added a \textit{network graph}, which shows the \textit{cross-loadings} (i.e., manifest variables with large loadings on more than one factor\footnote{\textit{Sparsity}, which partially refers to the absence of cross-loadings, is often taken as a proxy for interpretability, described in \textcite{Browne2001MBR}. A manifest variable should have a high loading for only one factor (i.e., no cross-loadings), which translates to the network graph as non-overlapping networks.}). A stacked bar chart is also available to help relate a theory to the model (i.e., adding \textit{a priori} tags to the manifest variables and looking at how much of each tag contributes to each factor). A key feature highlighted by \textcite{favis} is the capability to change and select the optimal threshold for dichotomizing the loading matrix. This interactivity is certainly important for performing EFA as one is able to quickly interpret the model as the threshold changes and ultimately find a configuration that is ideal.

\begin{figure}[H]
  \centering
  \begin{subfigure}{.5\textwidth}
      \centering
      \caption{Poor Interpretability}
      \includegraphics[width=1\textwidth]{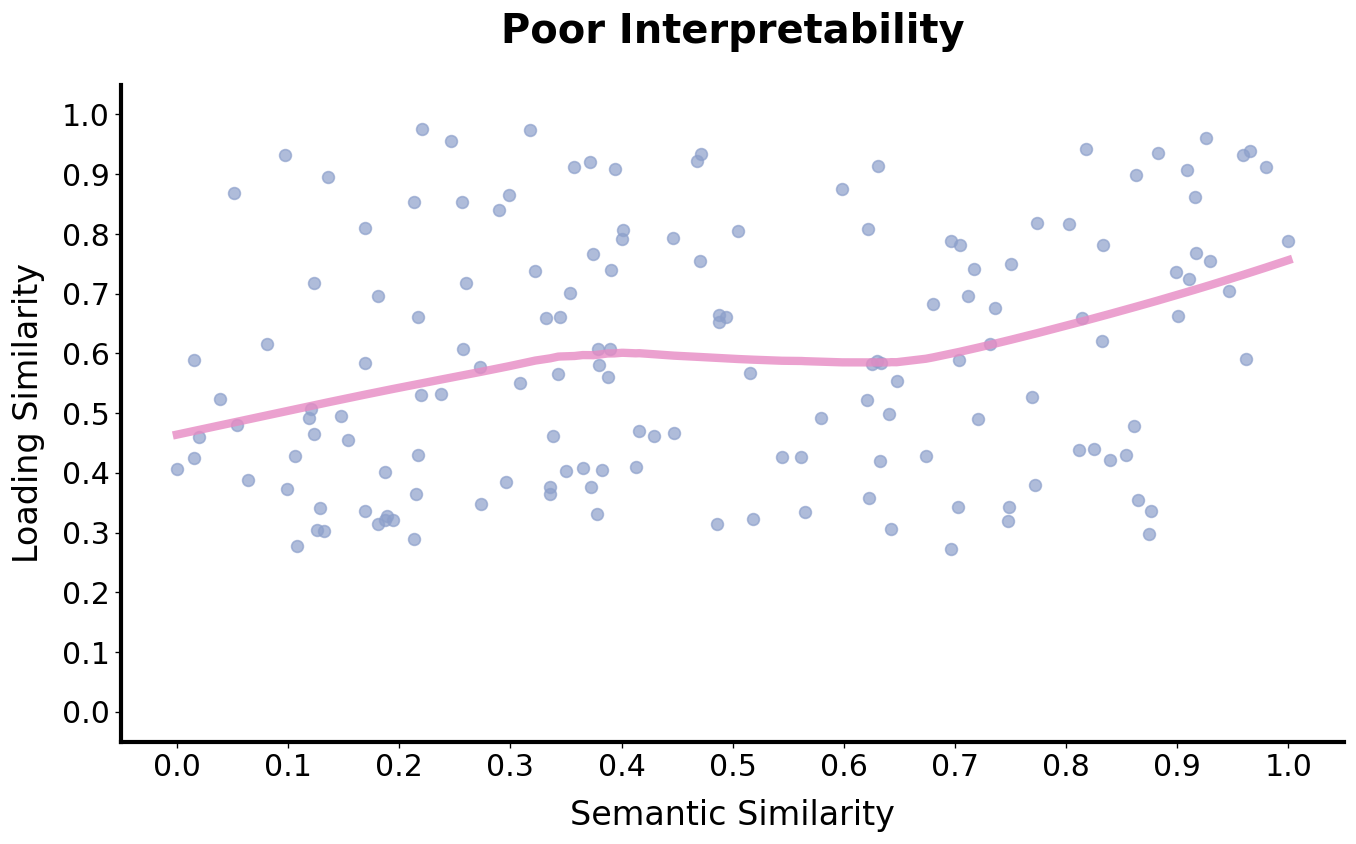}
  \end{subfigure}%
  \begin{subfigure}{.5\textwidth}
      \centering
      \caption{Good Interpretability}
      \includegraphics[width=1\textwidth]{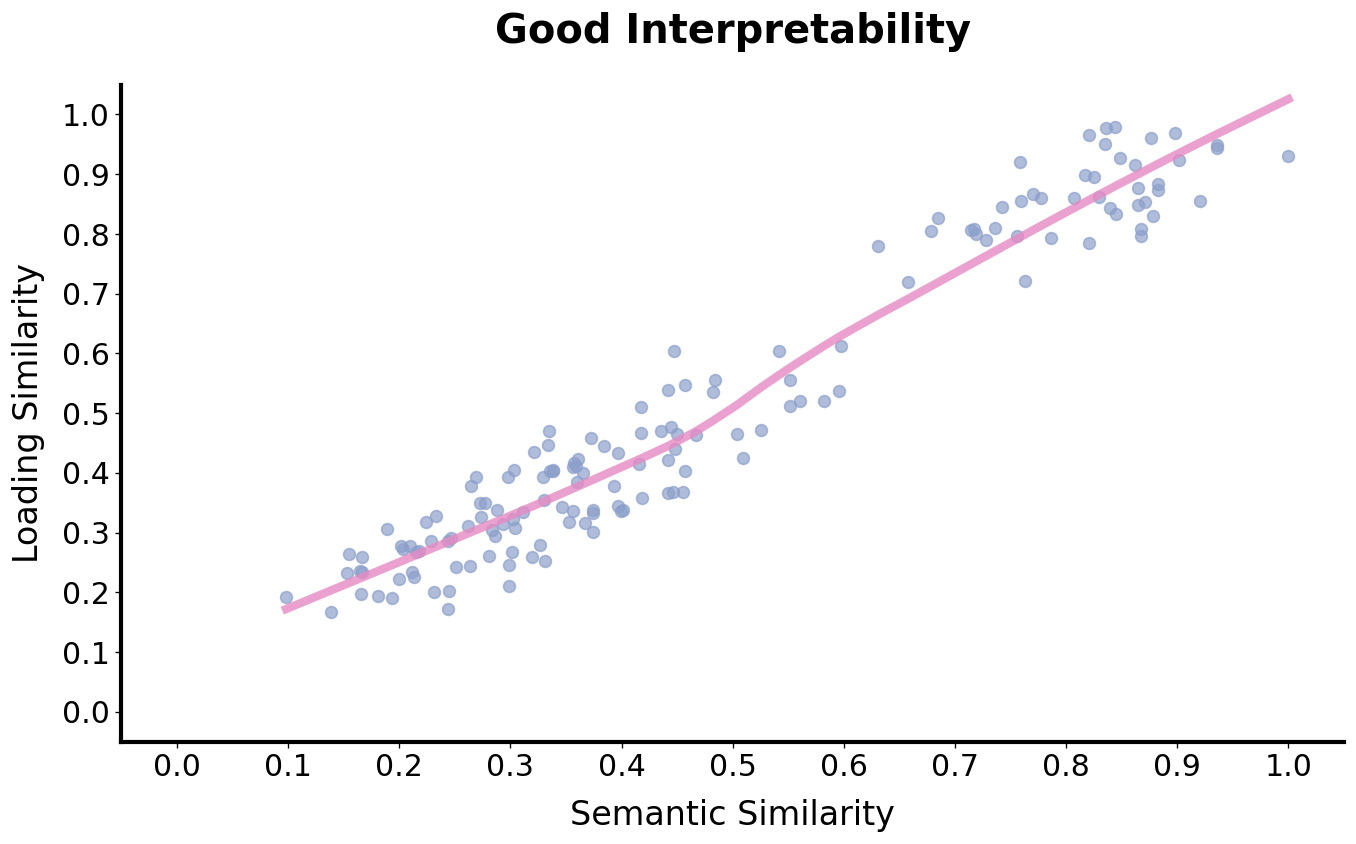}
  \end{subfigure}
  \caption{Interpretability Plots}
  \label{fig:example-interp}
\end{figure}

Finally, there are also nontraditional or new visualizations for EFA. One example is the \textit{interpretability plot} proposed by \textcite{tuazon2026}. The interpretability plot directly visualizes the interpretability (at least, a definition of it) of a factor model by showing how much the loadings align with the semantics of the questions (or in general, with \textit{a priori} information or expectations). Examples of the interpretability plot are shown in Figure \ref{fig:example-interp}. It is a scatter plot that shows the relationship between the \textit{loading similarity} and the \textit{semantic similarity} (or prior similarity). Ideally, the overall trend should be monotonically increasing because such indicates that the loadings ``agree" with the meanings of the questions (in the case of a questionnaire), or with the \textit{a priori} information or expectations (when using the more general prior similarities), which then corresponds to a meaningful and semantically coherent factor structure. Essentially, in the case of questionnaires, the model is interpretable if the ``clustering" of manifest variables based on the loading matrix is consistent with the ``clustering" of items based on semantic similarities, yielding semantically coherent factors.

\section{FactorFlow}\label{sec:ff}
\subsection{Overview}
As discussed previously, visualizations are important tools when it comes to performing EFA and due to the exploratory and iterative nature of EFA, dashboards can be even more beneficial. Here, we explore several areas where improvements can be made:
\begin{enumerate}
    \item As noted by \textcite{favis}, FAVis does not support visualizing and directly evaluating multiple factor models at the same time. Because of the exploratory nature of EFA, it is often the case that the researcher needs to compare different models and rotations to find the ``ideal" one, as explained earlier, which can be challenging to implement since large amounts of information must be shown for each model to adequately support interpretation.
    \item With the rise in the popularity and availability of powerful larage language models, it is arguably worthwhile to explore how large language models can be integrated into the EFA workflow. For instance, it can possibly be used to generated automated factor interpretations, making the process of interpreting a model much easier.
    \item Instead of focusing exclusively on providing visual analytics for interpreting the factor model, we aim to cover to end-to-end workflow for EFA. For example, we consider visualizations and figures for supporting exploratory data analysis and diagnostics (e.g., evaluating communalities, deciding on the number of factors, basic statistics), and a mechanism for fitting factor models with different configurations.
    \item The set of visualizations available in the dashboard can potentially be expanded. For instance, additional toggles or filters can be included, and even new visualizations (such as the interpretability plot, which was discussed earlier) can be added.
\end{enumerate}

Thus, this paper introduces \textit{FactorFlow}, a novel visual analytics workspace with large language model-assisted interpretation for (exploratory) factor analysis, with the following key features:
\begin{enumerate}
    \item Supports core exploratory factor analysis, implementing classical estimation methods, rotations, and visualizations, including various correlation types
    \item Provides implementations of pairwise target rotation and interpretability plots from \textcite{tuazon2026} for going beyond the classical methods
    \item Integrates large language models to provide automated large language model-generated interpretations, written in natural language, to assist model interpretation
    \item Enables deep analysis workflows, with multiple tabs available for basic statistics, hypothesis testing, diagnostics, model comparisons, and exploratory analyses
\end{enumerate}

In addition to its general goal of enabling users to perform EFA end-to-end effectively, the system also aims to address the following specific goals, some of which were adopted and modified from \textcite{favis}:
\begin{enumerate}
  \item \textbf{G1 - Enable users to fit various factors models}. A user should be able to fit a factor model with customizable specifications based on their dataset. For example, factor models can differ in terms of estimation method, rotation method, number of factors, included manifest variables, and so on.
  \item \textbf{G2 - Support various correlation types}. Pearson correlation may not be the best type of correlation for ordinal data. Thus, there should also be an option to use polychoric correlation for ordinal-level variables.
  \item \textbf{G3 - Enable users to perform model diagnostics and explore raw data}. A user should be able to asses the goodness-of-fit of an estimated factor model, as well as the adequacy of the sample data for factor analysis. Likewise, general summary information about the raw data can be useful for tangent hypotheses.
  \item \textbf{G4 - Effectively present general information about fitted models}. Factor correlations, factor scores, and other information (e.g., correlation matrix) are important considerations when examining a factor model.
  \item \textbf{G5 - Effectively present useful information for factor interpretation}. Associations between manifest variables and factors, which are reflected in the loading matrix, are the primary figures used for interpreting a factor model.
  \item \textbf{G6 - Enable users to select a good threshold for binarizing factor loadings}. Sufficiently small loadings do not affect a factor's interpretation but can possibly make the process more difficult. Thus, binarizing (i.e., setting sufficiently small loadings to $0$) can be useful but there must be a balance between interpretability and model fit.
  \item \textbf{G7 - Provide information about the sparsity of a factor model}. Sparsity in the loading matrix is sometimes used as a proxy for interpretability in factor models.
  \item \textbf{G8 - Provide a way to encode an existing theory or \textit{a priori} information}. An existing theory can be used to attempt interpreting factors in a factor model, where one can gauge how much of each hypothesized construct is reflected in each factor. However, beyond classical rotation methods, \textit{a priori} information about the loading matrix can also be used to influence the loadings themselves.
  \item \textbf{G9 - Assist the user with interpreting factor models through generated natural language interpretations}. A large language model can be used to interpret a factor model based on the loading matrix and the associated statements.
  \item \textbf{G10 - Provide a mechanism to compare multiple models}. EFA generally involves comparing multiple candidate factor models. Thus, there should be a way to easily compare two models side-by-side.
  \item \textbf{G11 - Provide a comprehensive set of visualizations for assessing a factor model}. A factor model can be visualized in terms of the loadings, loading comparisons across manifest variables or factors, agreement with \textit{a priori} information or theory, sparsity, and goodness-of-fit.
\end{enumerate}

The current version of FactorFlow was developed using \codify{Python 3}, with ChatGPT from \textcite{openai2026chatgpt} used as a coding assistant tool. In particular, FactorFlow is a \textit{Streamlit} application and thus, uses the \codify{streamlit} package. Streamlit is an open-source Python framework for building data applications \parencite{st}. For interactive visualizations, the primary library used was \codify{plotly}. 

Custom \codify{JavaScript} was also written inside the Streamlit application to handle features related to natural language processing. The \codify{TensorFlow.js}\footnote{The universal sentence encoder can also be directly imported via \codify{Python} but due to limited resources in the cloud, the computations were offloaded to the front-end via \codify{JavaScript}.} was used to load the \textit{universal sentence encoder} from \textcite{use:2018} while the Groq application programming interface (API) was leveraged using the \codify{groq} Python package for generating automated interpretations.

The application's source code can be found on GitHub here: \url{https://github.com/jptuazon/factorflow}. As of writing, FactorFlow is hosted on the free tier of Streamlit cloud and can be accessed here: \url{https://factorflow-efa.streamlit.app/}.

\subsection{Design, Features, and Usage}
\subsubsection{General Parts}
The application has several parts. First, it has a \textbf{\textit{Menu}} \textbf{sidebar}. There are four different \textbf{panels} in the sidebar:
\begin{enumerate}
  \item \textbf{NLP Models}. This shows the statuses of the NLP models used in the app. You can also configure the large language model here.
  \item \textbf{Data}. This is where you can upload your datasets. You can also examine your dataset in this panel by clicking the Stats button. High-level information is also displayed here.
  \item \textbf{Factor Models}. You can fit new factor models using this panel. It also shows the factor models that you have fitted and saved so far. You can examine each saved models by clicking View. High-level information is also displayed here.
  \item \textbf{Settings}. You can configure the application to match your preferences. You can change things like the color palettes used and the chart styles.
\end{enumerate}

Moreover, it has several pages:
\begin{enumerate}
  \item \textbf{Overview}. This is the home page, where you can find general information about the app.
  \item \textbf{Diagnostics}. This is where you can examine diagnostics for factor analysis. Here, you can determine how many factors to use or which manifest variables have low communalities, and so on.
  \item \textbf{Dashboard}. In this tab, you can see multiple figures and visualizations for assessing the factor model. You can also compare two factor models at the same time here.
  \item \textbf{About}. You can find development details for the app here.
  \item \textbf{Privacy}. Information about how user data is handled can be found here.
\end{enumerate}

\begin{figure}[H]
    \centering
    \includegraphics[width=0.9\textwidth]{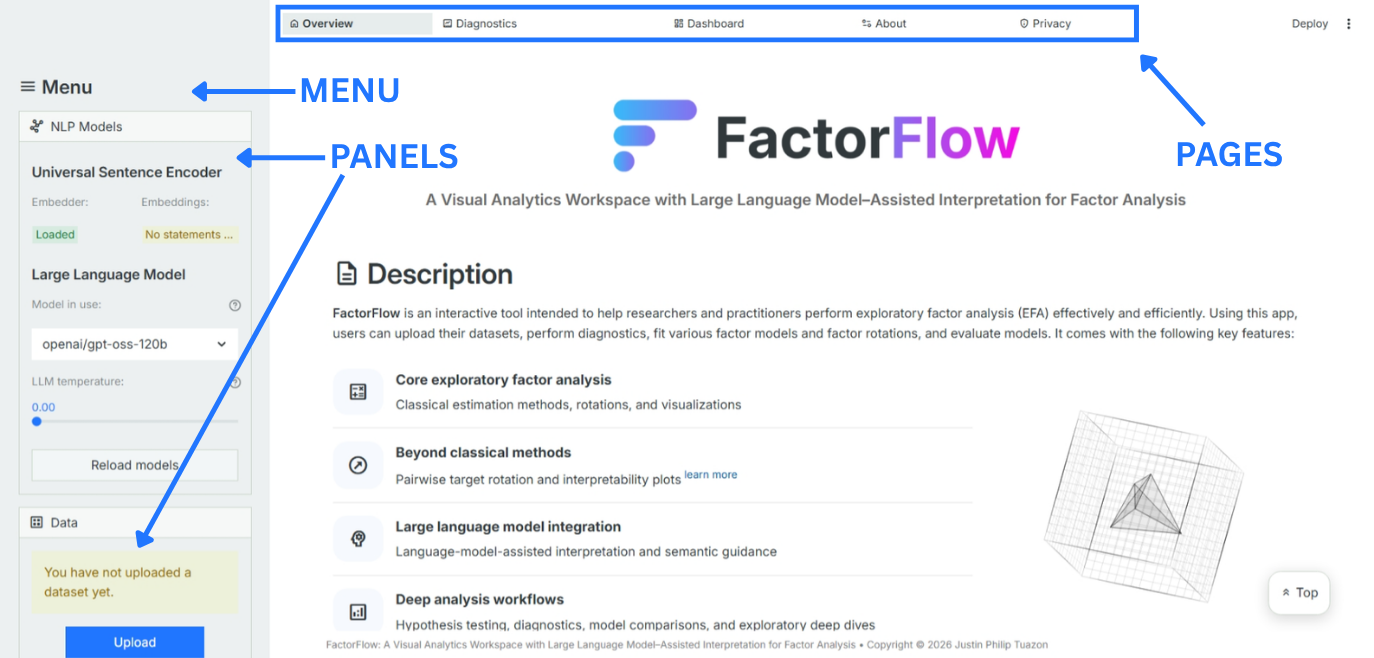}
    \caption{Parts of the App}
    \label{fig:parts-of-the-app}
\end{figure}

Generally, the application can handle \textbf{any tabular dataset}. However, the app works best when dealing with questionnaires that have \textbf{Likert-type items}. Some features are most useful only for Likert-type items.

\subsubsection{Menu}
\begin{figure}[H]
  \centering
  \begin{subfigure}[c]{1\textwidth}
      \centering
      \includegraphics[width=1\textwidth]{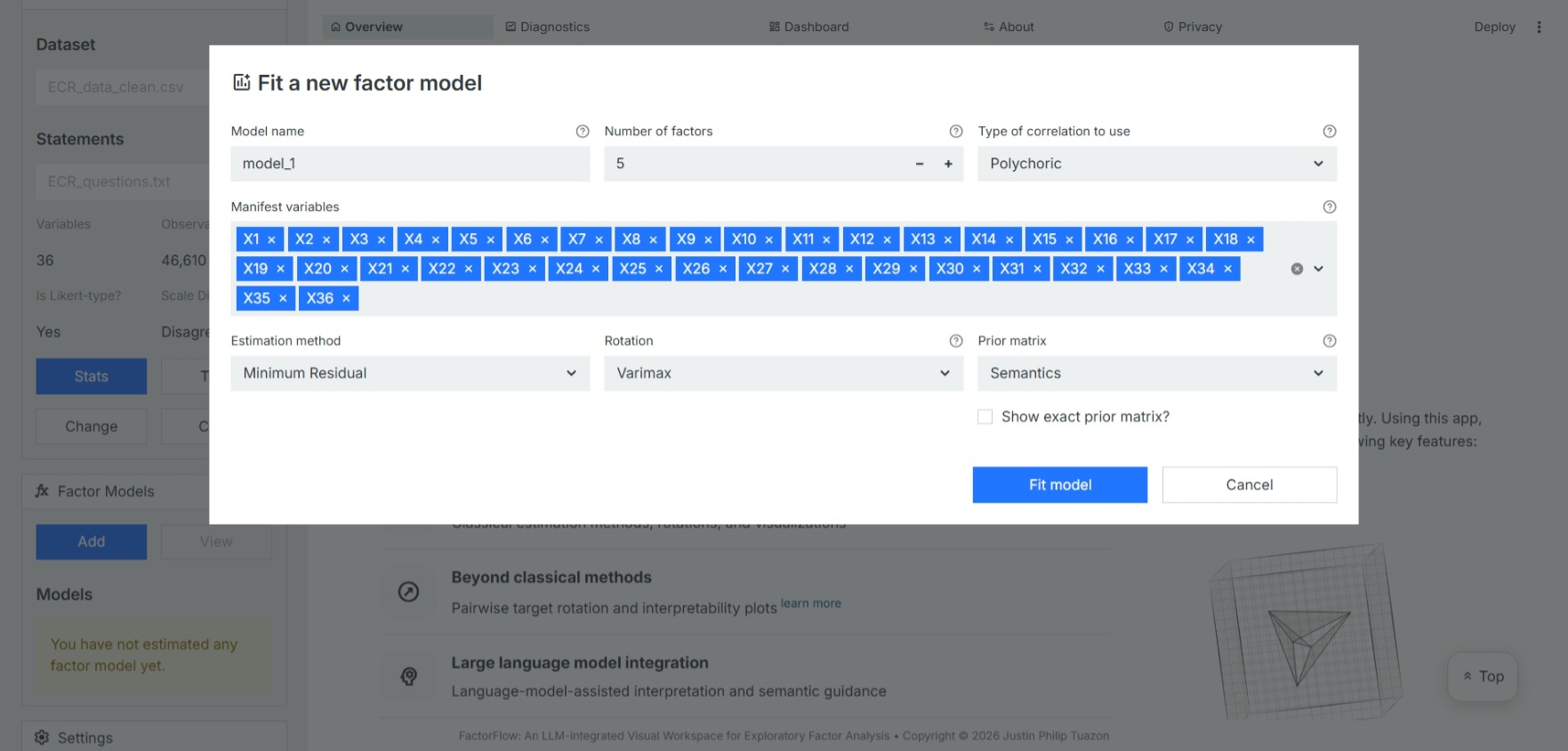}
      \caption{Fitting a Factor Model}
  \end{subfigure}
  \begin{subfigure}[c]{.5\textwidth}
      \centering
      \includegraphics[width=1\textwidth,]{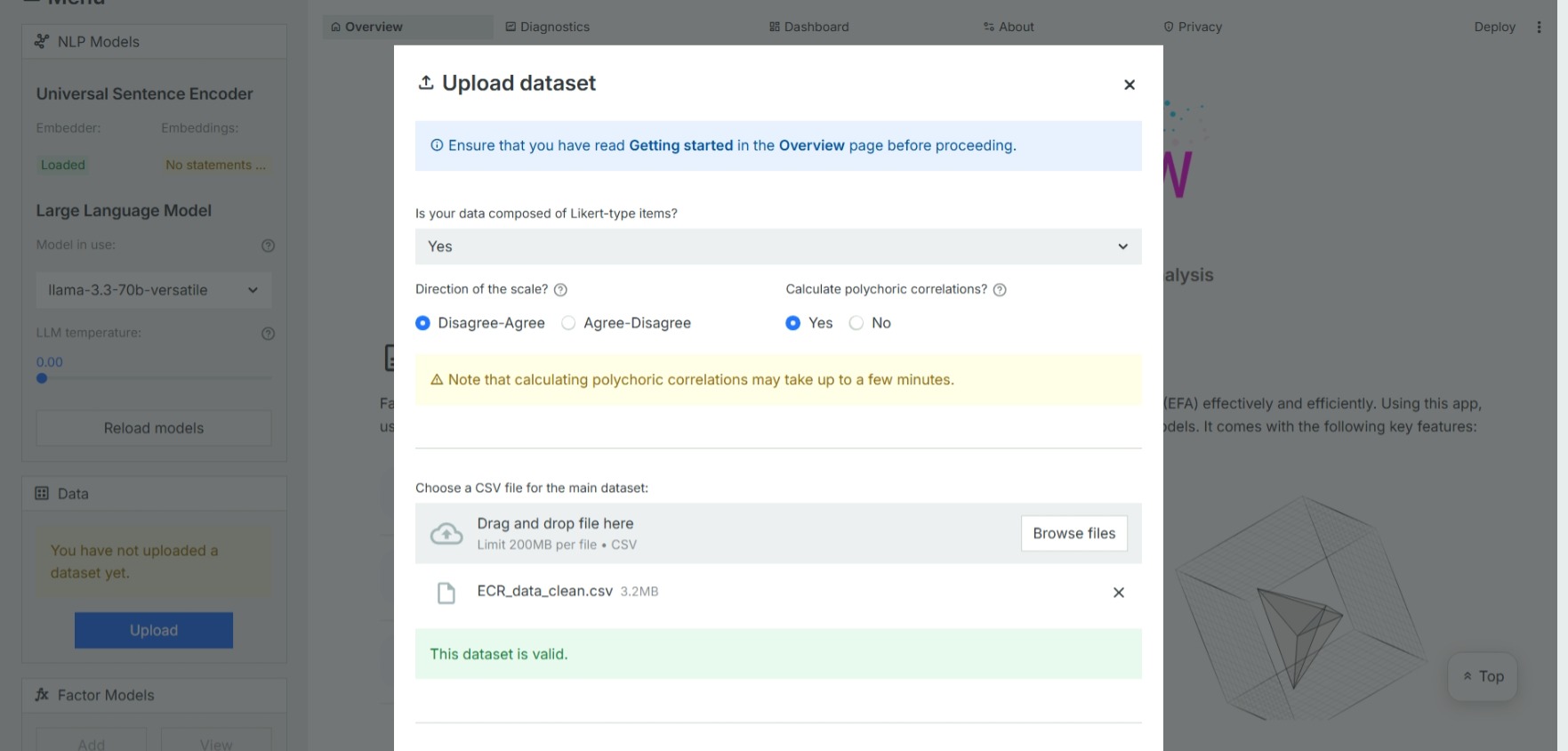}
      \caption{Uploading a Dataset\protect}
  \end{subfigure}%
  \begin{subfigure}[c]{.5\textwidth}
      \centering
      \includegraphics[width=1\textwidth]{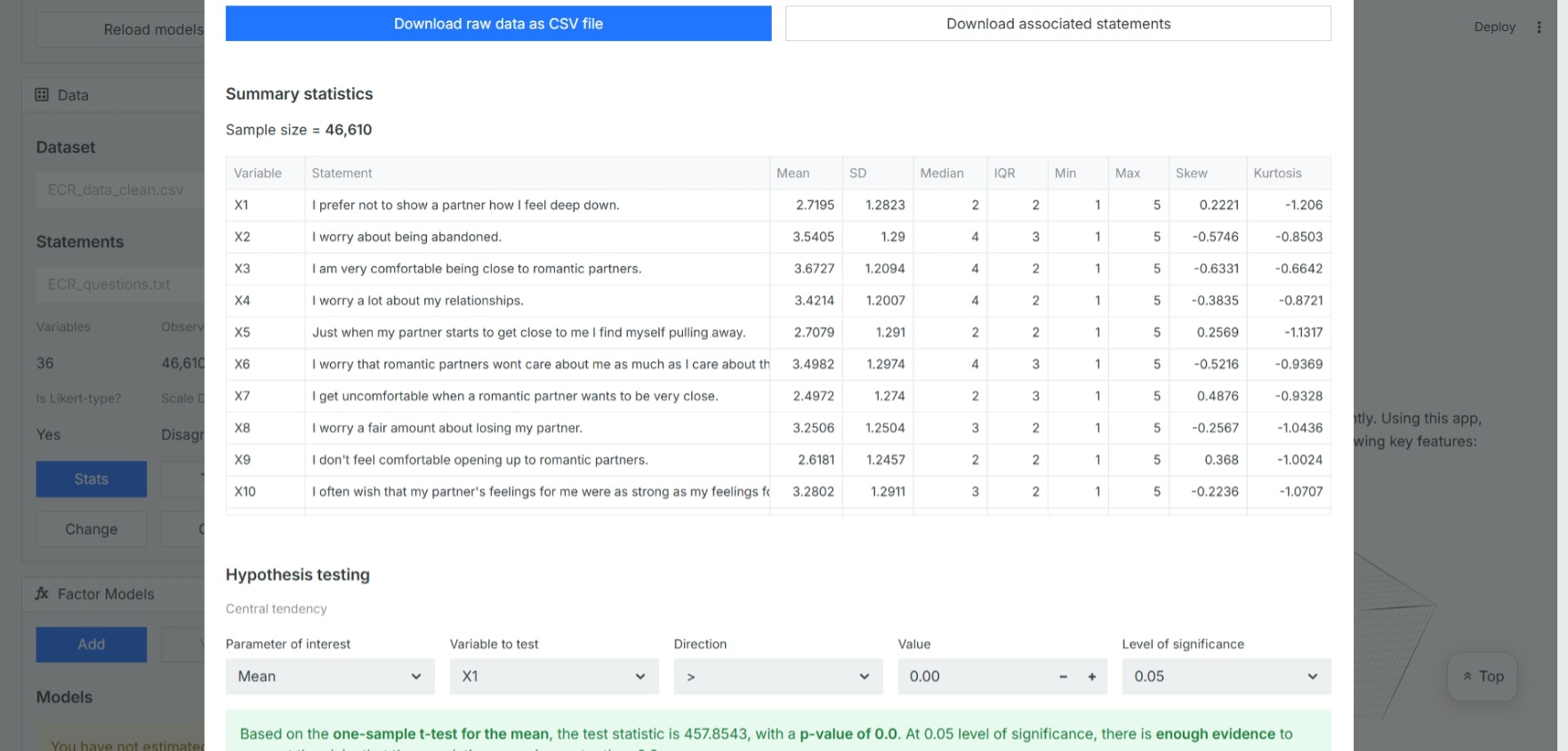}
      \caption{Basic Statistics\protect}
  \end{subfigure}
  \caption{Primary Menu Actions}
  \label{fig:primary-menu-actions}
\end{figure}

As mentioned before, the sidebar has several panels. The first useful panel is likely the \textbf{\textit{Data}} panel, where the user can upload their dataset. When a dataset has been uploaded, the user can perform basic exploratory data analysis and hypothesis testing by clicking the \textbf{\textit{Stats}} button under the \textbf{\textit{Data}} panel. Under the \textbf{\textit{Factor Models}} panel, one can add or view estimated factor models. Screenshots of the app showing examples of these features can found in Figure \ref{fig:primary-menu-actions}. Collectively, the actions available in the sidebar address goals \textbf{G1}, \textbf{G2}, and \textbf{G4}.

\subsubsection{Diagnostics}
\begin{figure}[H]
    \centering
    \includegraphics[width=0.5\textwidth]{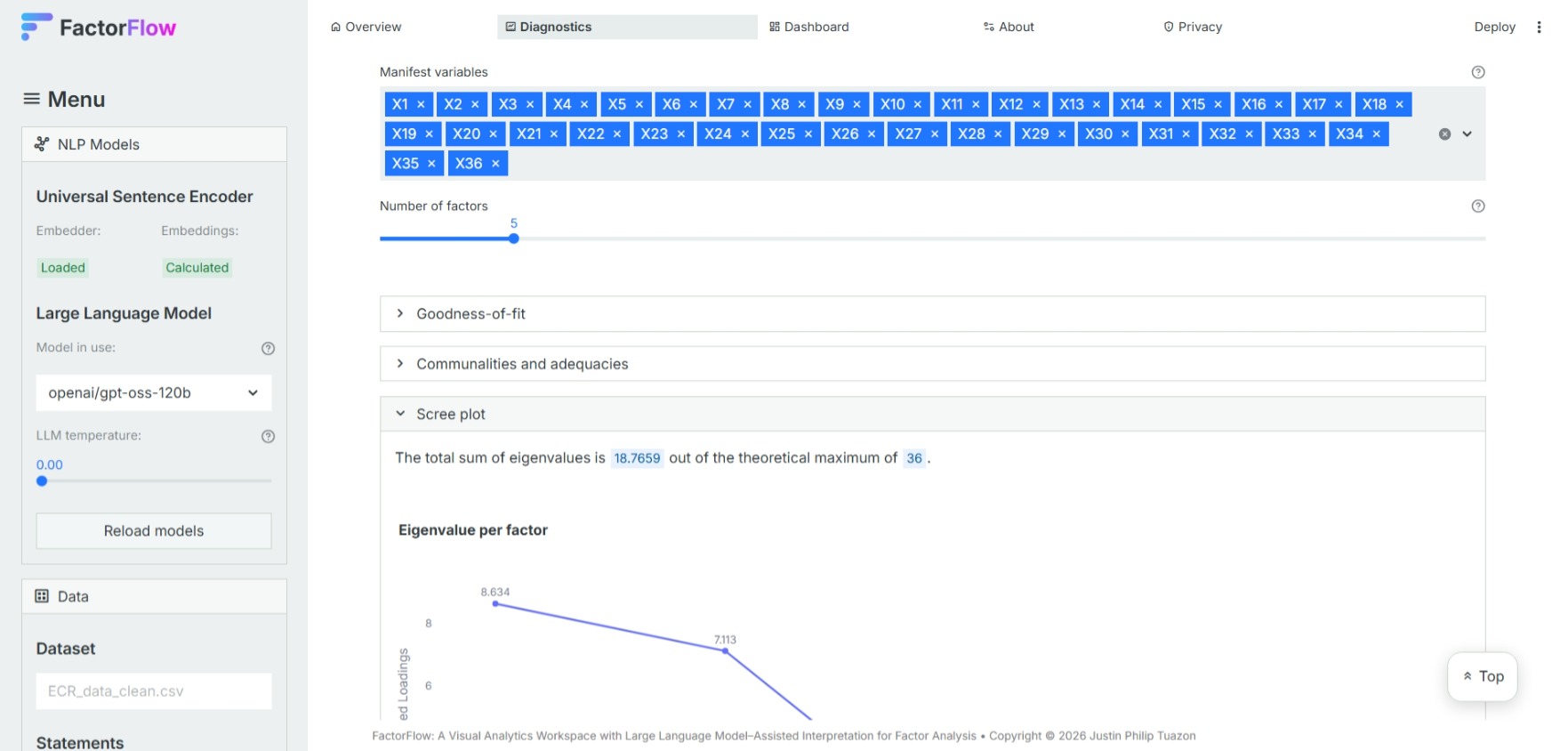}
    \caption{Diagnostics Page}
    \label{fig:diagnostics}
\end{figure}

The \textbf{\textit{Diagnostics}} page, a portion of which is shown in Figure \ref{fig:diagnostics}, has filters for the \textit{manifest variables} and the \textit{number of factors}. This addresses goal \textbf{G3}. It has four different sections, whose outputs depend on the selected values in the filters:
\begin{enumerate}
  \item \textbf{Goodness-of-fit}. Several goodness-of-fit indices (e.g., root mean square error of approximation) can be found here.
  \item \textbf{Communalities and adequacies}. This section shows information about the communality and the sampling adequacy of each manifest variable in the model. One can use this to identify which manifest variables should be excluded from the model.
  \item \textbf{Scree plot}. This shows a scree plot that can help the user identify how many number of factors to consider in the model.
  \item \textbf{Correlations}. The correlation matrix for the manifest variables can be found here. It has an additional filter where the user can select the type of correlation matrix to display (i.e., Pearson or polychoric).
\end{enumerate}

\subsubsection{Dashboard}
The \textbf{\textit{Dashboard}} is perhaps the most important page in FactorFlow for assessing factor models. It has a filter for selecting which models (up to two) to analyze. Then, information about the selected models are shown in the dashboard, which has eight sections\footnote{Each bullet is annotated with the main goal(s) that it addresses.}:
\begin{enumerate}
  \item \textbf{Fit details}\textsuperscript{\textbf{G4}}. This shows information about the model's fit details (e.g., rotation, correlation type).
  \item \textbf{Communalities and adequacies}\textsuperscript{\textbf{G4}}. This shows the communality and adequacy for each manifest variable included in the factor model, informing the user about how much the factor model explains each manifest variable.
  \item \textbf{Interpretability plot}\textsuperscript{\textbf{G5},\textbf{G8}}. This shows the scatterplot, with a locally weighted scatterplot smoothing (LOWESS) curve, on semantic (or prior) similarity vs loading similarity. This visualization can tell you how much the loading similarities agree with the semantic (or prior) similarities. The V-index is displayed below the title of the plot (higher value is better).
  \item \textbf{Factor breakdown}\textsuperscript{\textbf{G5},\textbf{G8}}. This shows how much of each tag each factor is composed of, in terms of sum of squared loadings. This can help you interpret the factors based on the tags that you have provided (if any). For example, if Factor 1's sum of squared loadings comes mostly from Tag A, then you can associate the meaning of Factor 1 with Tag A.
  \item \textbf{Factor loadings}\textsuperscript{\textbf{G5},\textbf{G6}}. This shows you the empirical distribution (in terms of the cumulative distribution function) of the absolute loadings, both for each factor and for the whole model. You can use the distribution plots to asses which threshold for the absolute loadings make sense. Below the distribution plot, the factor loading matrix, presented as a heatmap, can be seen. You can choose whether to show the exact loadings or the dichotomized loadings (i.e., the loading is replaced with $-1$ if the loading is negative and has a large magnitude, $1$ if the loading is positive and has a large magnitude, and $0$ otherwise). This will help you define what each factor is (and assess whether the factors make sense based on the definitions).
  \item \textbf{Loadings comparison}\textsuperscript{\textbf{G5}}. The plot here allows you to compare the loadings across manifest variables or across factors. For example, when comparing across factors, if two lines are coincident or parallel, then the corresponding factors are similar (in terms of definition).
  \item \textbf{Factor cross-loadings}\textsuperscript{\textbf{G5},\textbf{G7}}. This network graph tells you how much the factors overlap. For example, if the ``network" of Factor 1 shares multiple nodes with the ``network" of Factor 2, there are multiple cross-loadings (i.e., significant overlap) between the two factors.
  \item \textbf{Interpretation}\textsuperscript{\textbf{G9}}. You can generate automated interpretations for factor models (at a factor-level) here using the selected large language model.
\end{enumerate}

\begin{figure}[H]
\centering
\begin{subfigure}[c]{0.5\textwidth}
  \centering
  \includegraphics[width=1\textwidth]{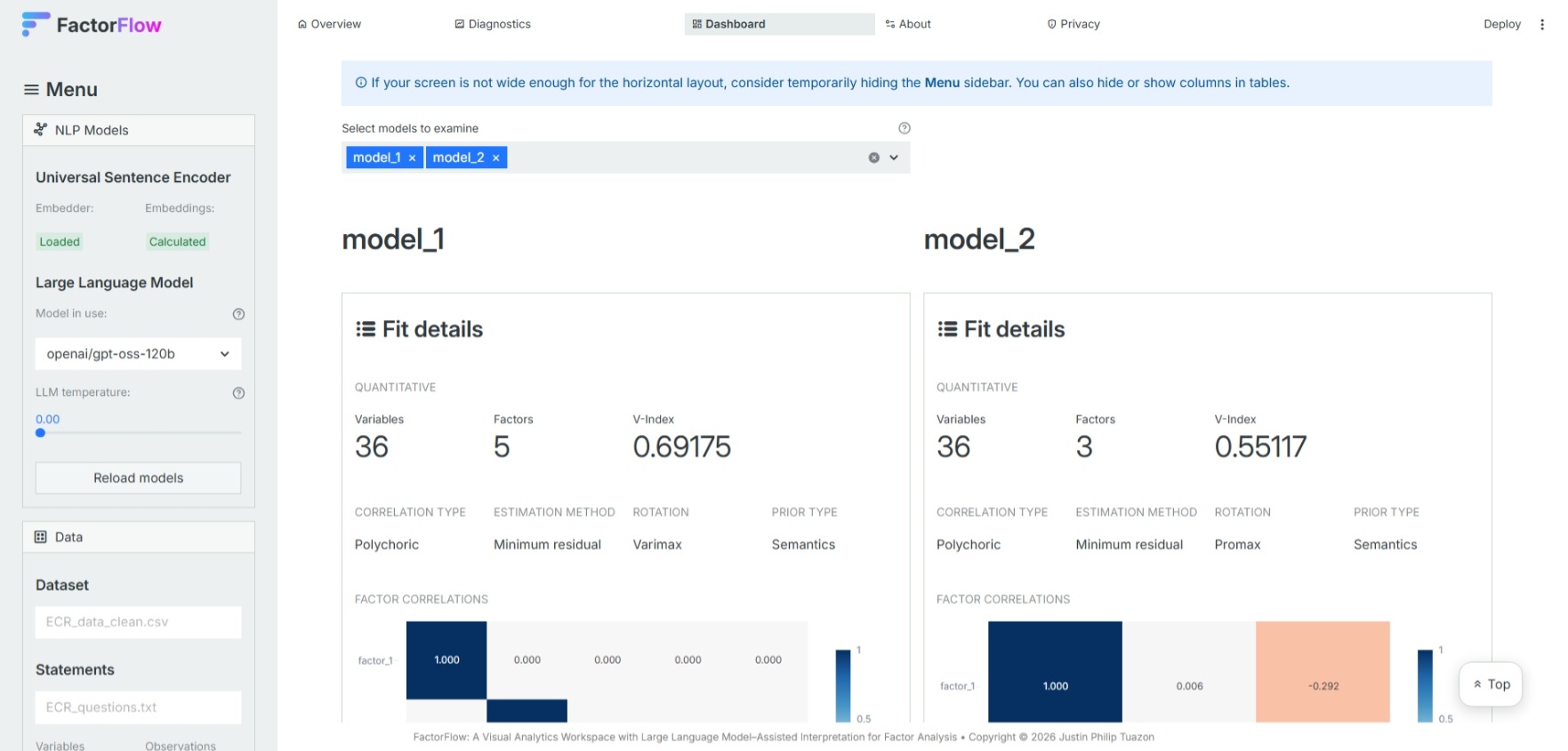}
  \caption{Fit Details}
\end{subfigure}%
\begin{subfigure}[c]{0.5\textwidth}
  \centering
  \includegraphics[width=1\textwidth]{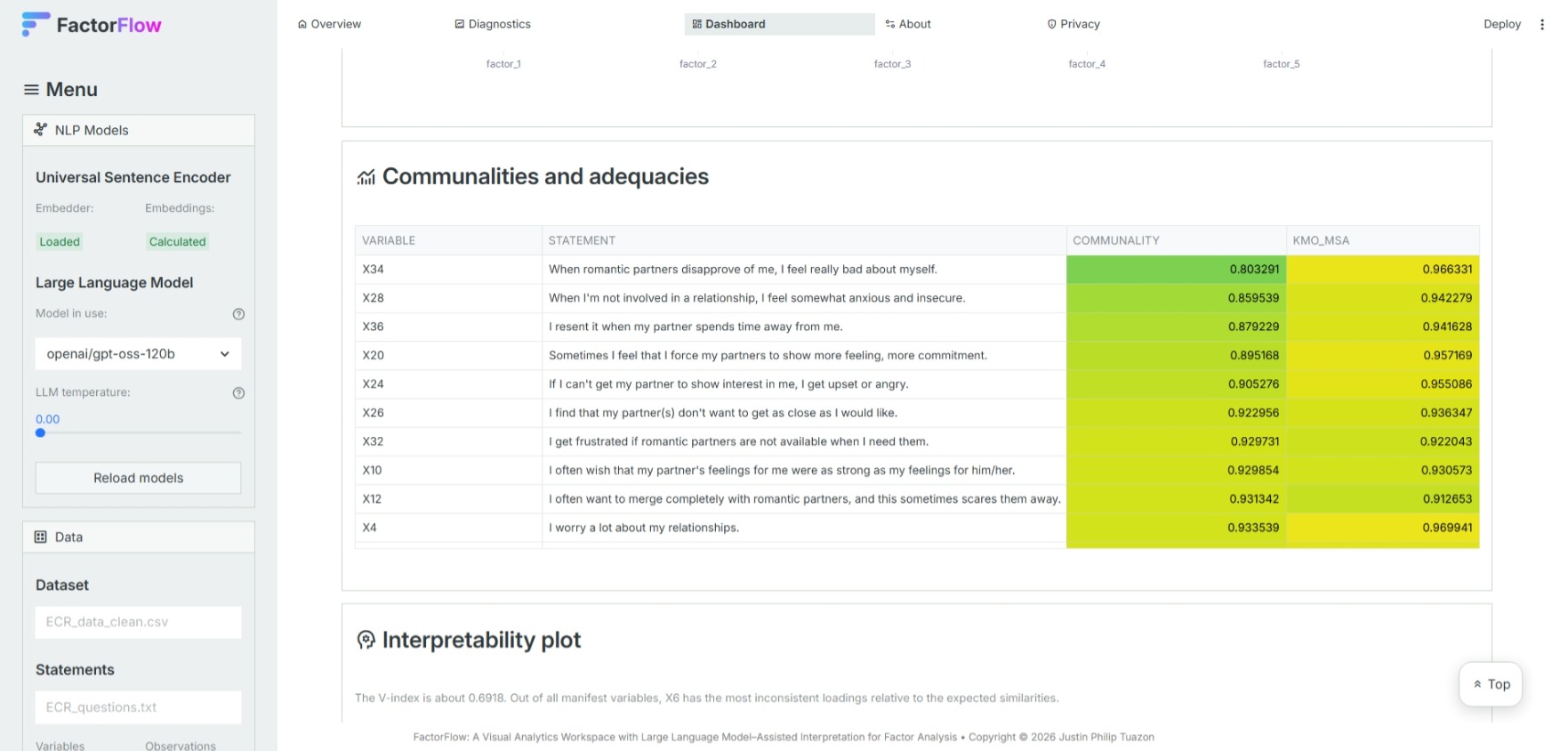}
  \caption{Communalities and Sampling Adequacies}
\end{subfigure}

\begin{subfigure}[c]{0.5\textwidth}
  \centering
  \includegraphics[width=1\textwidth]{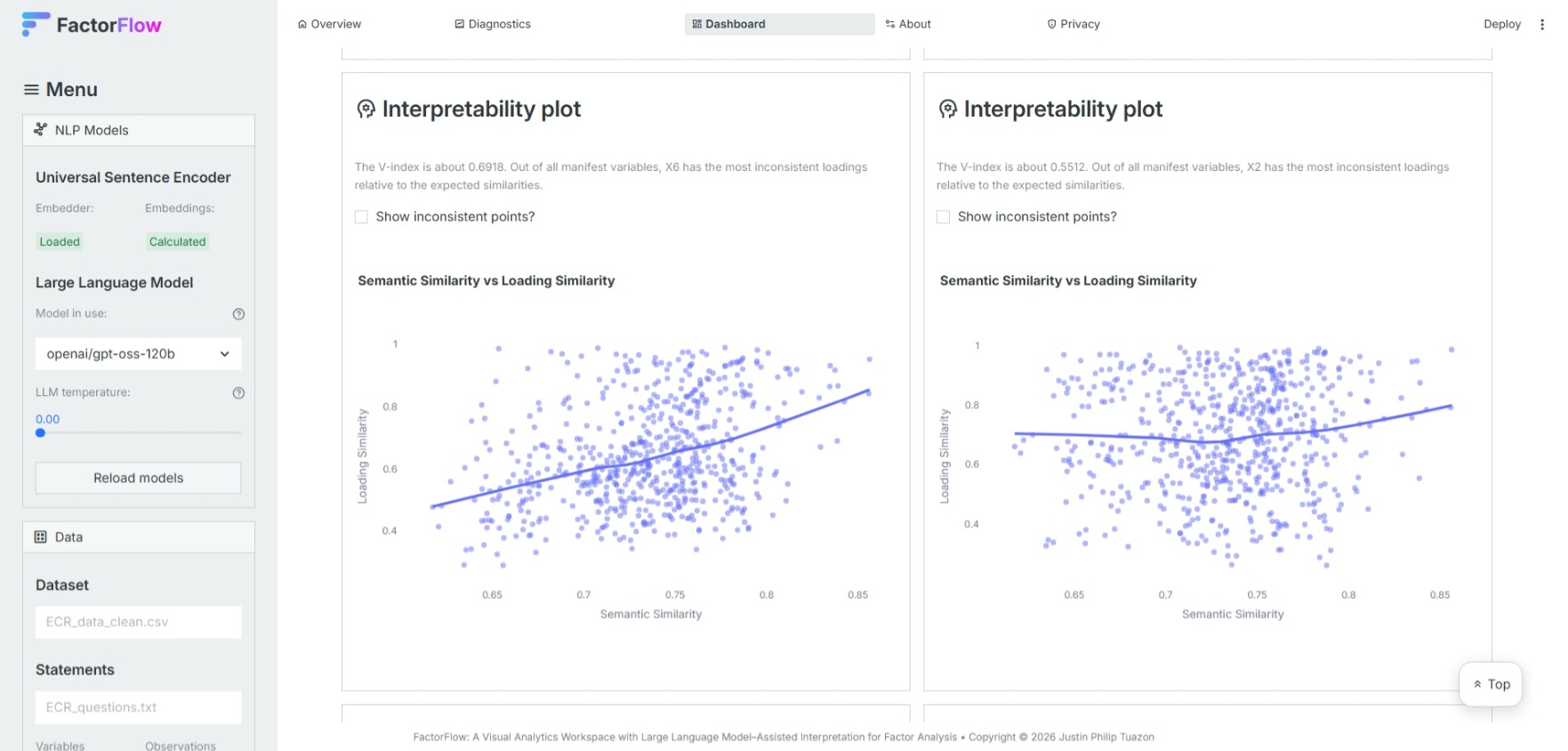}
  \caption{Interpretability Plot}
\end{subfigure}%
\begin{subfigure}[c]{0.5\textwidth}
  \centering
  \includegraphics[width=1\textwidth]{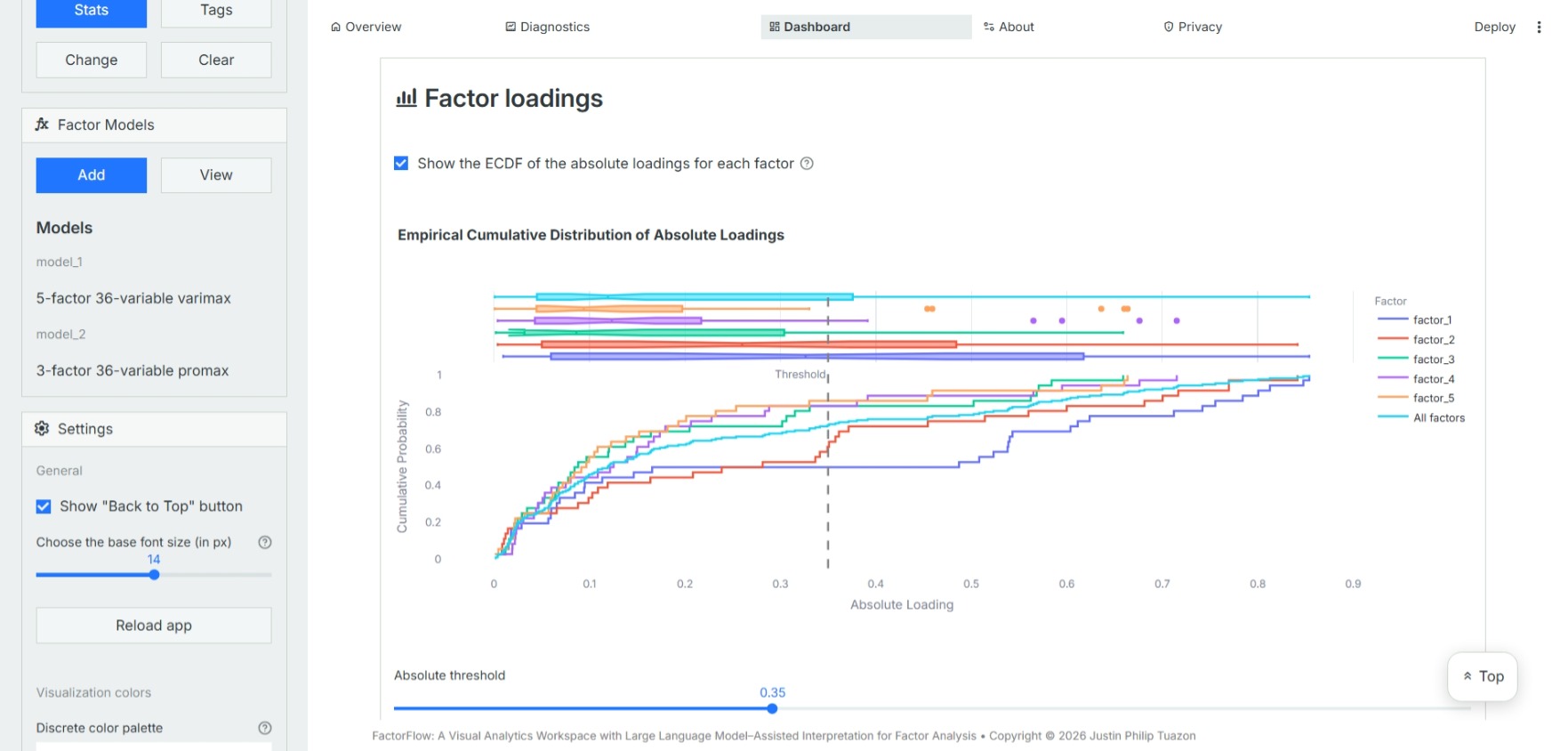}
  \caption{Loadings Distribution}
\end{subfigure}

\begin{subfigure}[c]{0.5\textwidth}
  \centering
  \includegraphics[width=1\textwidth]{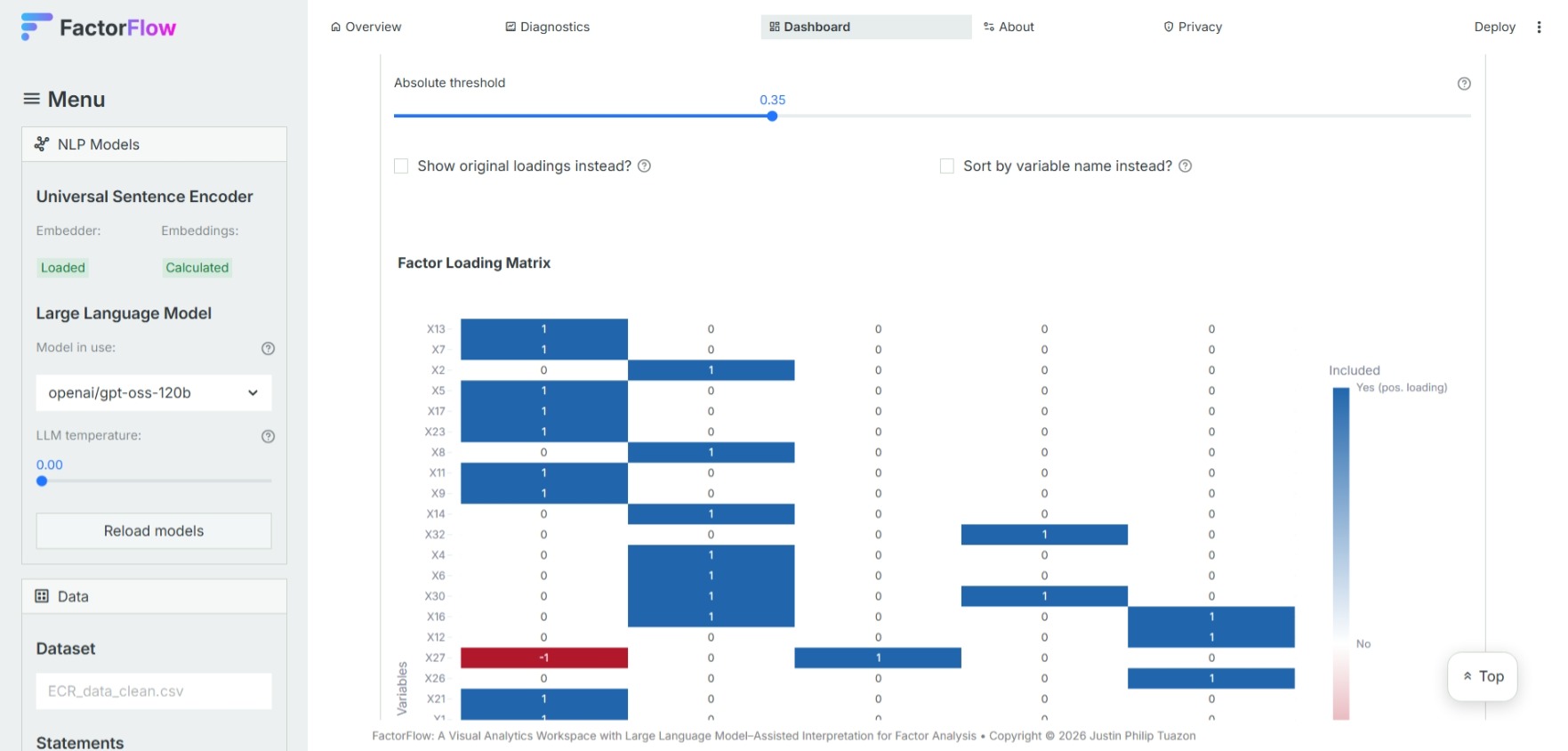}
  \caption{Loading Matrix}
\end{subfigure}%
\begin{subfigure}[c]{0.5\textwidth}
  \centering
  \includegraphics[width=1\textwidth]{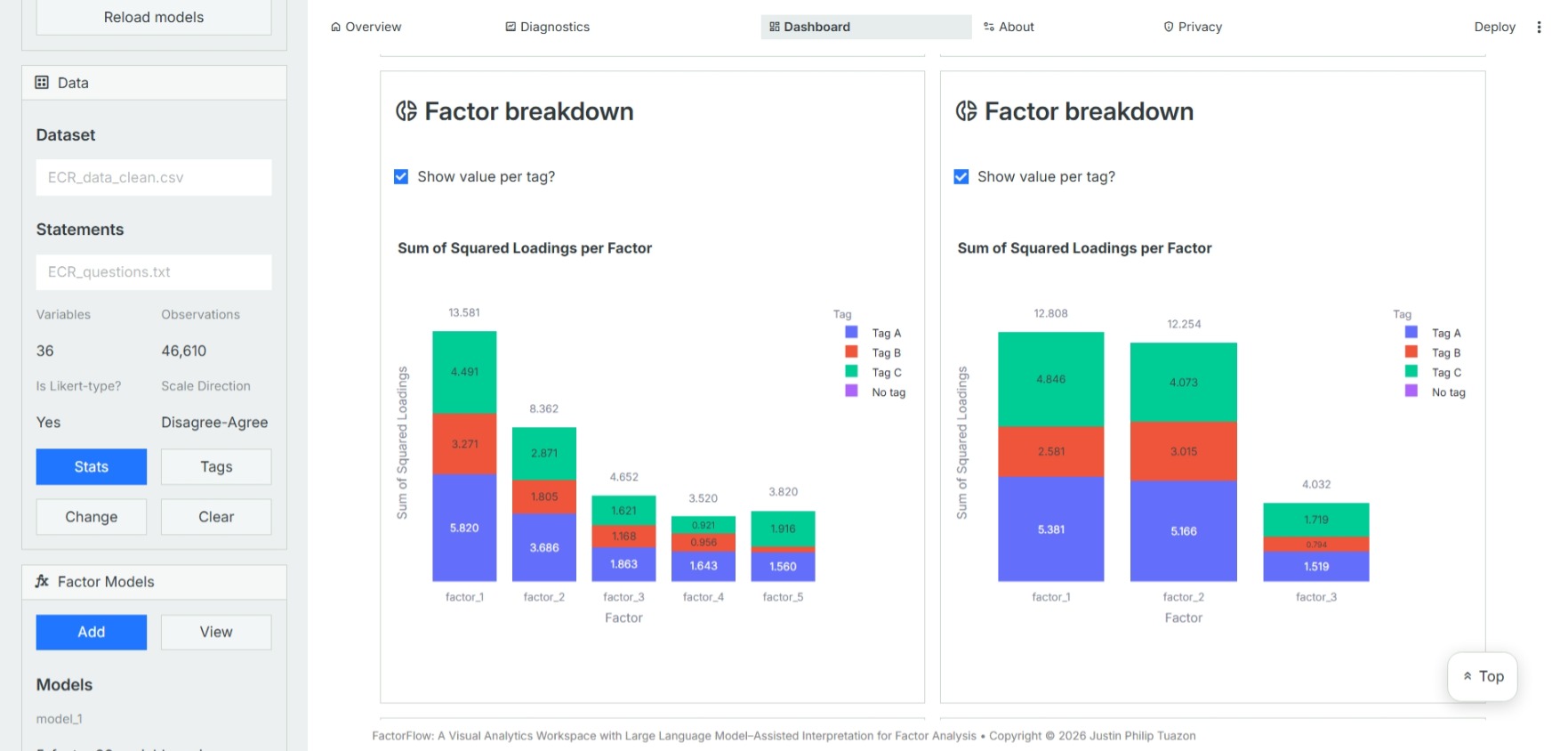}
  \caption{Factor Breakdown}
\end{subfigure}

\begin{subfigure}[c]{0.5\textwidth}
  \centering
  \includegraphics[width=1\textwidth]{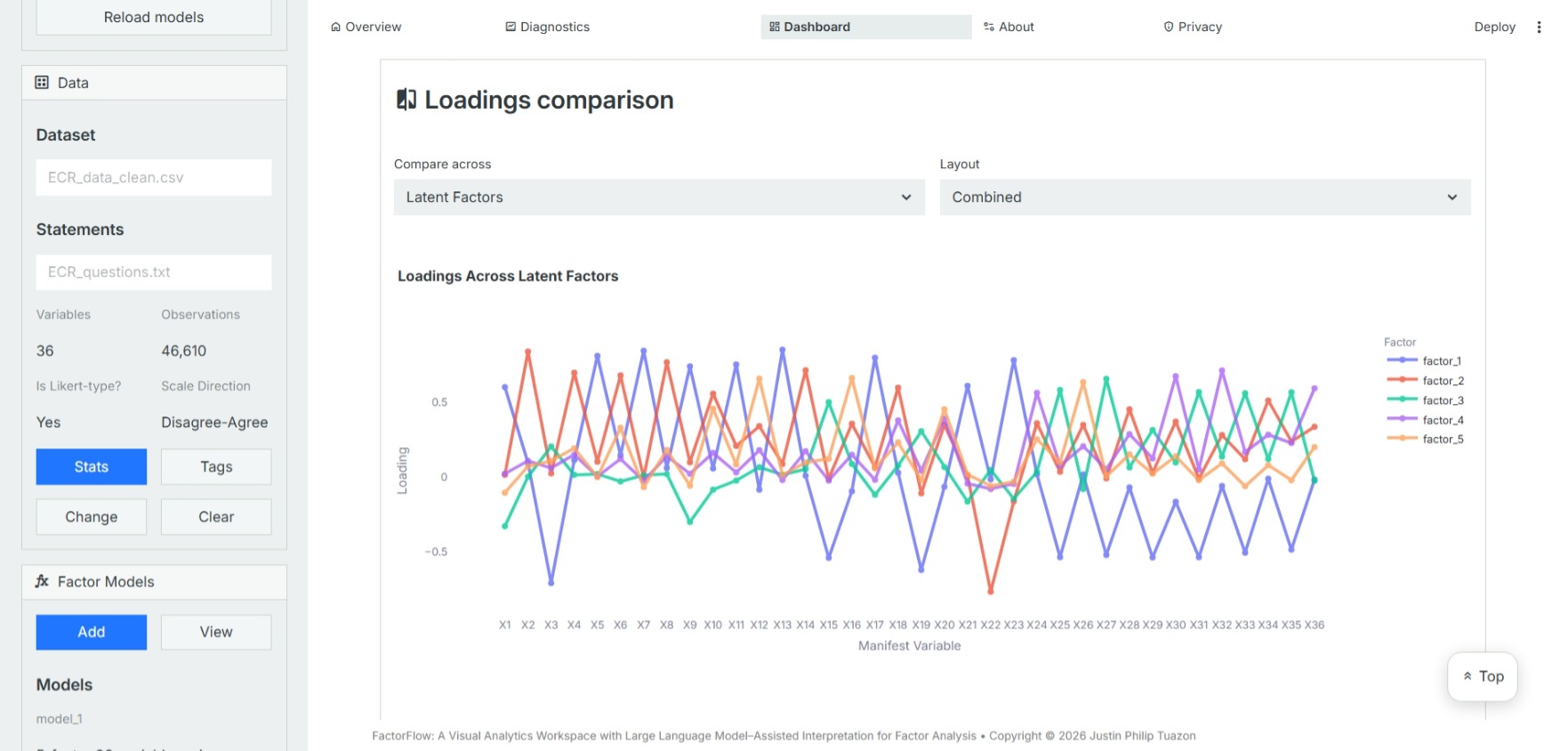}
  \caption{Loadings Comparison}
\end{subfigure}%
\begin{subfigure}[c]{0.5\textwidth}
  \centering
  \includegraphics[width=1\textwidth]{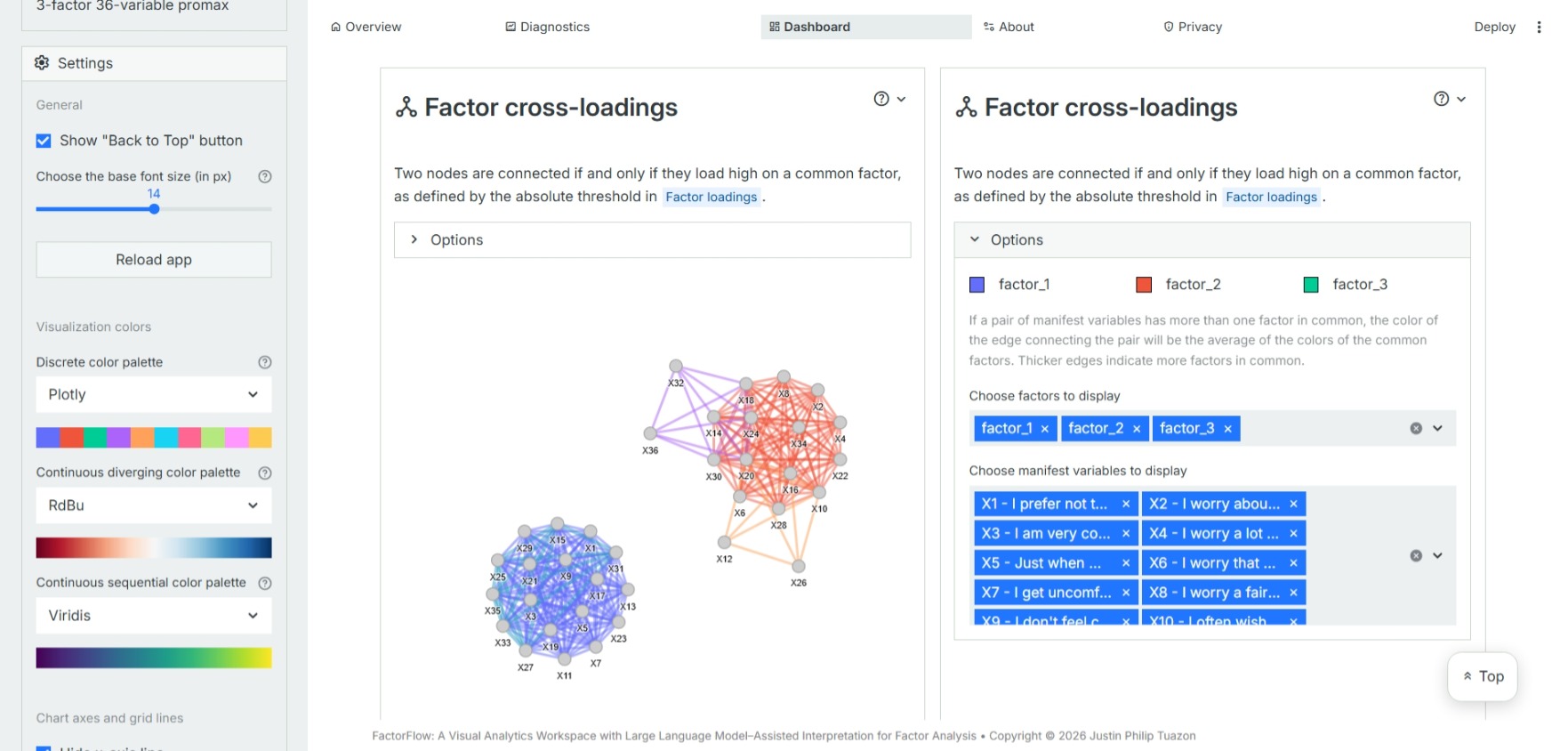}
  \caption{Factor Cross-loadings}
\end{subfigure}

\begin{subfigure}[c]{0.5\textwidth}
  \centering
  \includegraphics[width=1\textwidth]{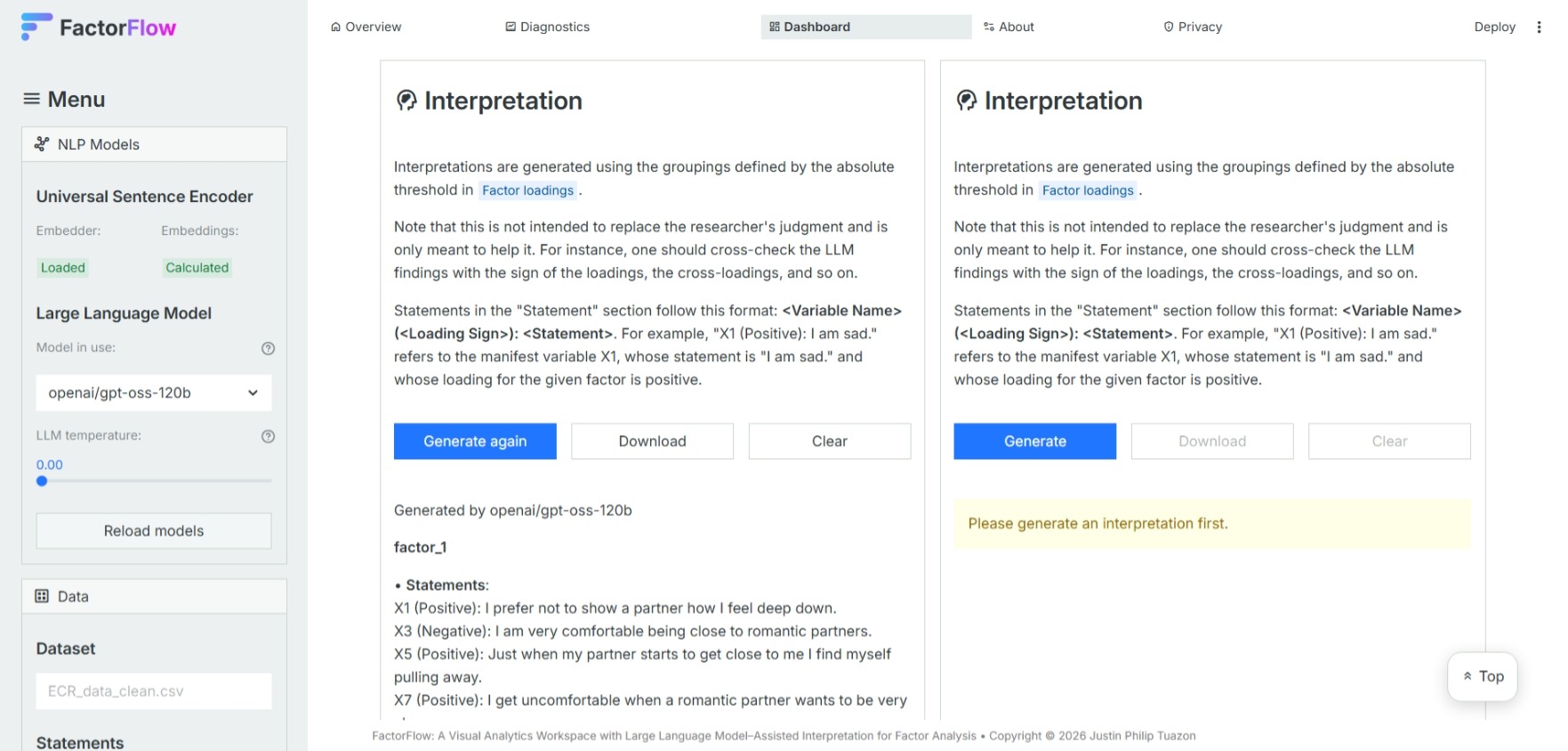}
  \caption{Generating Automated Interpretations}
\end{subfigure}%
\begin{subfigure}[c]{0.5\textwidth}
  \centering
  \includegraphics[width=1\textwidth]{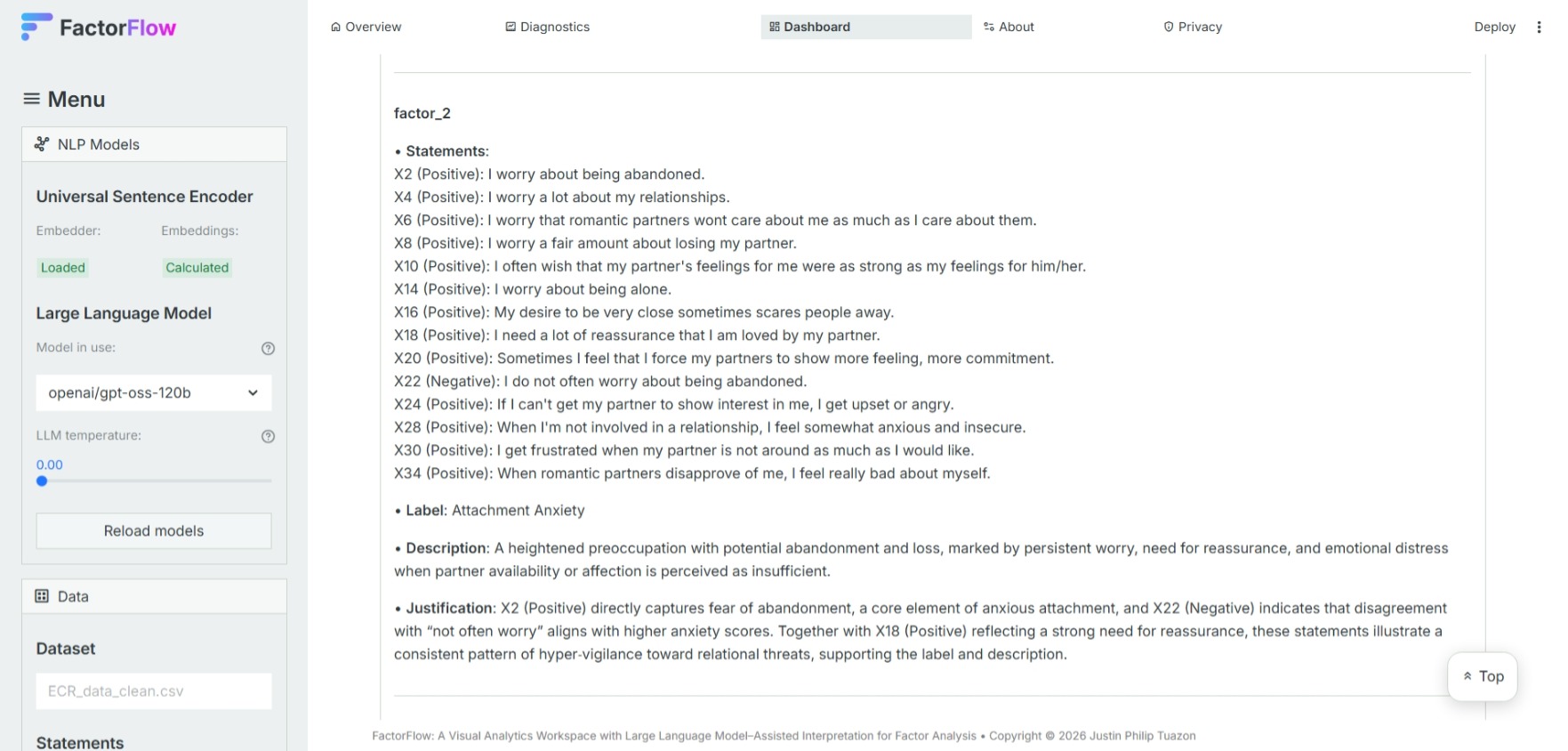}
  \caption{Sample Automated Interpretation}
\end{subfigure}

\caption{Some Parts of the \textit{Dashboard} Page}
\label{fig:dashboard}
\end{figure}

Each section, sample screenshots of which are shown in Figure \ref{fig:dashboard}, provides information about a different aspect of the factor model. \textbf{Fit details} is useful for general information, while \textbf{Communalities and adequacies} provide goodness-of-fit measures. \textbf{Interpretability plot} and \textbf{Factor breakdown} are useful for comparing the model with \textit{a priori} information (through pairwise similarities) or theory (through individual variable-level tags), and on the other hand, \textbf{Factor loadings} and \textbf{Interpretation} facilitate the actual interpretation of individual factors. Finally, \textbf{Loadings comparison} and \textbf{Factor cross-loadings} assist with global interpretation (i.e., the entire factor model instead of individual factors) by providing information about redundancy and about sparsity, respectively. The different sections in \textbf{\textit{Dashboard}} cover various goals: \textbf{G4}, \textbf{G5}, \textbf{G6}, \textbf{G7}, \textbf{G8}, and \textbf{G9}. Collectively, the dashboard also addresses \textbf{G10} and \textbf{G11}.

Note that, as hinted earlier, a user can easily compare two models side-by-side. To make such possible, a parallel vertical layout was chosen, where each model is assigned a column and corresponding figures for a given model are shown from top to bottom, as opposed to a mosaic layout similar to \textcite{favis}. Columns are then placed side-by-side with corresponding dashboard sections  aligned across the columns for ease of comparison. If only one model is chosen, one column spanning the entire width is allocated for the single model.

\subsubsection{Automated Interpretation}
\begin{figure}[H]
    \centering
    \includegraphics[width=0.8\textwidth]{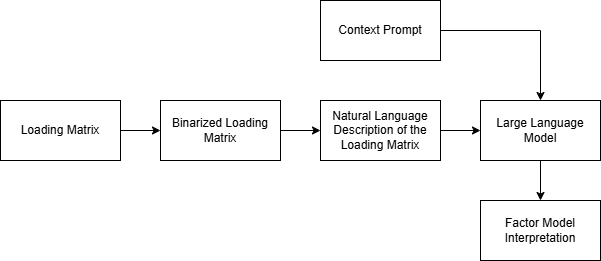}
    \caption{Generating Automated Interpretations}
    \label{fig:llm-flow}
\end{figure}

Figure \ref{fig:llm-flow} shows the high-level summary of how a large language model is utilized to produce automated factor interpretations.

When generating automated factor interpretations using large language models, the context for the language model is first set using the prompt shown in Appendix \ref{appendix:A}. In general, the context prompt does the following things:
\begin{enumerate}
    \item The general role (i.e., an ``EFA Factor Interpretation Assistant") and task (i.e., interpret each factor) are given.
    \item Specific critical instructions, primarily those that ensure the completeness of the output and correctness of the format, are provided.
    \item Requirements for each major section (i.e., Statements, Description, and Justification) are then set (e.g., ``avoid surface-level or generic interpretations").
\end{enumerate}
Moreover, aside from describing the formats of the input and output, the context prompt also sets several rules:
\begin{enumerate}
    \item \textbf{Scale Direction Rule}. This explains how scale direction (e.g., Disagree to Agree) should affect interpretation.
    \item \textbf{Conflict Handling Rule}. This provides the procedure to follow when the factor appears to be incoherent.
    \item \textbf{Empty Factor Rule}. This provides instructions on what to output when the factor has no associated statements.
    \item \textbf{Ambiguity Handling Rule}. This explains how ambiguous statements should be pointed out.
    \item \textbf{Redundancy Rule}. This ensures that sections do not repeat the same pieces of information.
    \item \textbf{Formatting Rules}. This explains how the output must be formatted.
\end{enumerate}
Finally, output requirements are emphasized to ensure that the final output of the large language model completes the task adequately. Note that in the 
\textbf{\textit{NLP Models}} panel, the user can set the \textit{temperature}, as well as the large language model in use, to decide how to generate the automated interpretation (e.g., more creative, more deterministic).

Now, when constructing the input for the large language model, the loading matrix is first processed as follows to ultimately produce an appropriate description of the matrix in natural language:
\begin{enumerate}
    \item The direction of the scale (e.g., higher values indicate higher degrees of agreement) is noted.
    \item Each factor is associated with a ``group" of manifest variables. For every factor, only the manifest variables whose absolute loadings meet or exceed the threshold is included in the corresponding group.
    \item Under each factor, every manifest variable is converted into this format: ``$<$Variable Name$>$ ($<$Loading Sign$>$): $<$Statement$>$". For example, ``$X1$ (Positive): I am sad." refers to the manifest variable $X1$, whose statement is ``I am sad." and whose loading for the given factor is positive.
    \item The overall converted loading matrix (i.e., the natural language description of the loading matrix) is then constructed, an example of which is shown in Appendix \ref{appendix:B}.
\end{enumerate}

After inputting the context prompt and the converted loading matrix prompt, the large language model then returns an interpretation of the model in the following block format, with one block for each factor:
\begin{Verbatim}[breaklines]
    factor_<factor number>
    • Statements:
    <manifest variable name> (<sign of loading>): <corresponding statement>
    <manifest variable name> (<sign of loading>): <corresponding statement>
    <manifest variable name> (<sign of loading>): <corresponding statement>
    • Label: <a 2-4 word label for the factor>
    • Description: <an explanation of the latent construct>
    • Justification: <the reasoning behind the interpretation of the factor>
\end{Verbatim}

\subsubsection{User Personas and Sample Workflow}
In general, the application is geared towards \textbf{any researcher or practitioner who needs to perform EFA} and the recommended workflow is as follows:
\begin{enumerate}
  \item \textbf{Configure}. This is an optional step but the user can set preferences or stick with the defaults (e.g., color palettes, chart styles) under the \textbf{\textit{Settings}} panel.
  \item \textbf{Upload}. The user should upload their dataset(s) under the \textbf{\textit{Data}} panel.
  \item \textbf{Explore}. The user should examine basic statistics on the raw data and perform some model diagnostics, by clicking the \textbf{\textit{Stats}} button under the \textbf{\textit{Data}} panel and going to the \textbf{\textit{Diagnostics}} tab.
  \item \textbf{Fit}. The user can then fit various factor models by clicking \textbf{\textit{Add}} under the \textbf{\textit{Factor Models}} panel.
  \item \textbf{Analyze}. The user can then head on to the \textbf{\textit{Dashboard}} tab to analyze one or more models.
  \item \textbf{Export}. Finally, the user can download desired visualizations and other outputs.
\end{enumerate}

However, \textbf{any user} can also just utilize the \textbf{\textit{Stats}} button or feature (under \textbf{\textit{Data}} panel) if they just wish to examine the raw data and perform basic hypothesis tests on it. For example, FactorFlow is more useful for \textit{constructing} questionnaires (e.g., scales) compared to \textit{summarizing} responses, but it can be used for both.

For a detailed and step-by-step guide (i.e., a sample workflow), the user can click \textbf{\textit{Want more detailed instructions?}} under the \textbf{\textit{Getting started}} section of the \textbf{\textit{Overview}} page.

\begin{figure}[H]
  \centering
  \begin{subfigure}[c]{0.5\textwidth}
      \centering
      \includegraphics[width=1\textwidth]{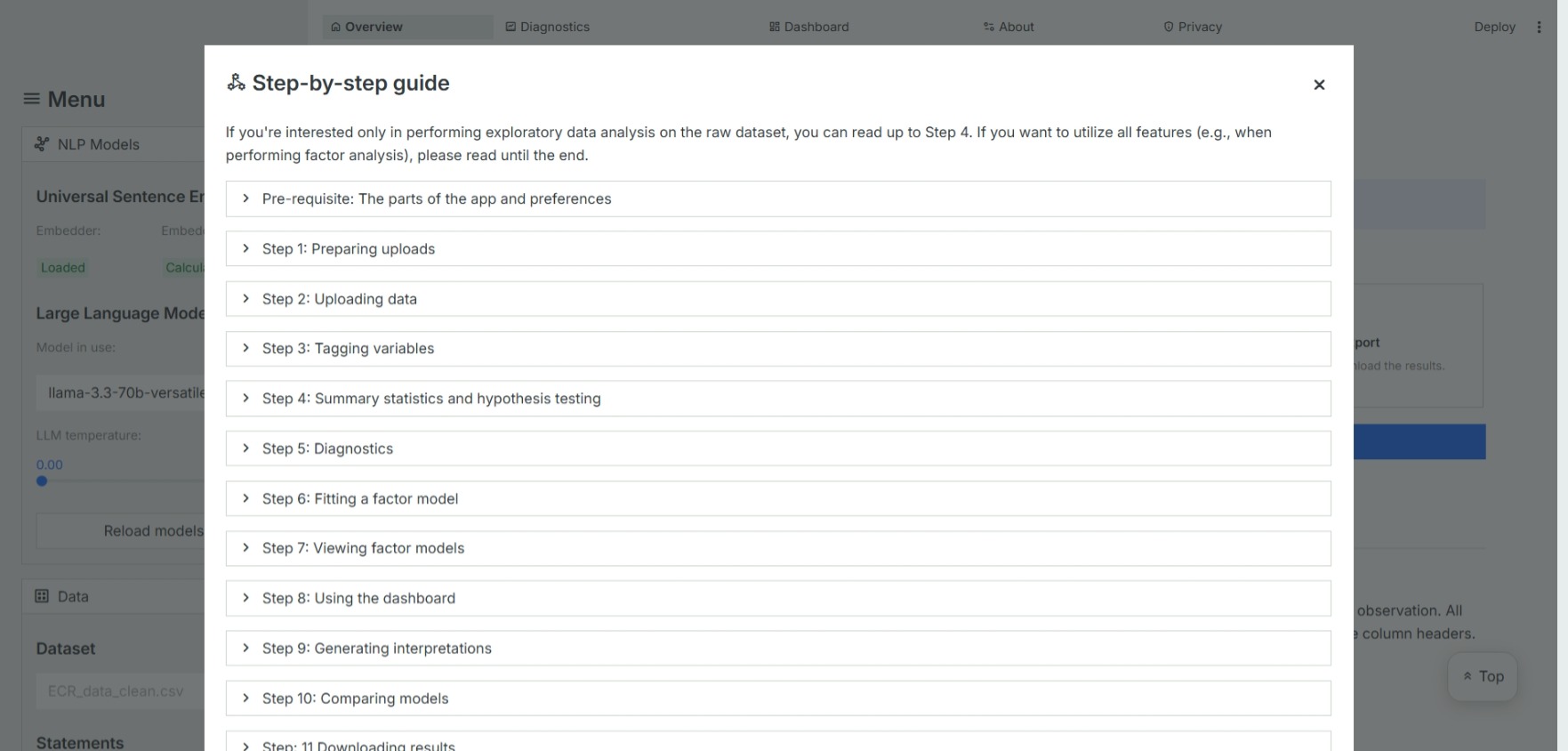}
  \end{subfigure}%
  \begin{subfigure}[c]{0.5\textwidth}
      \centering
      \includegraphics[width=1\textwidth]{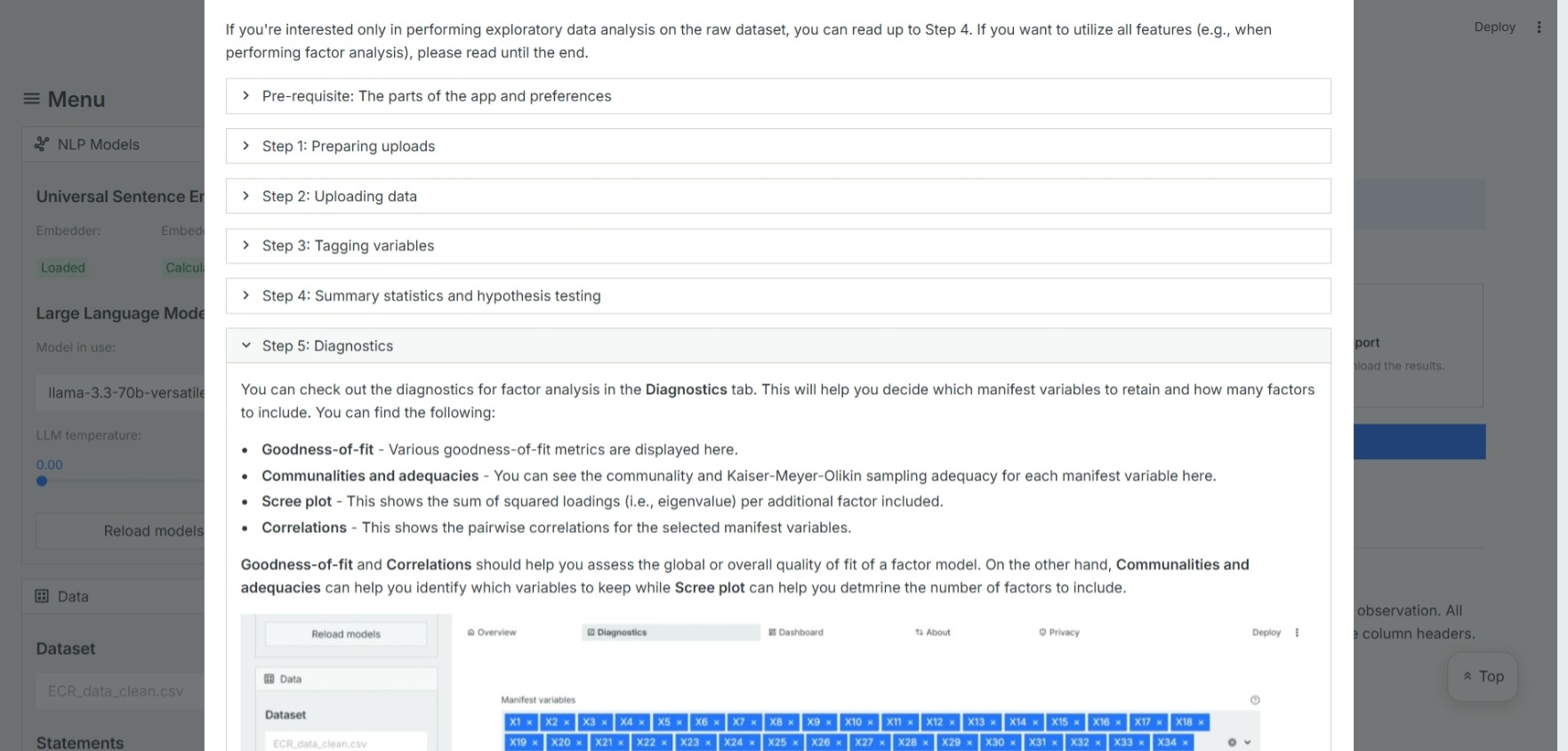}
  \end{subfigure}
  \label{fig:guide}
  \caption{Step-by-step Guide or Sample Workflow for FactorFlow}
\end{figure}

\subsection{Usability Survey}
A usability study was conducted to gather feedback about the application. Similar to \textcite{favis}, the survey was informal in the sense that it had only $5$ participants, who were all selected via convenience sampling. Still, the survey was extremely useful as its primary goal was to gather feedback and not conduct statistical inference.

The survey consisted of several sections. One gathered information about the respondent's demographics, which also ensured that the respondent had experience with EFA. Then, other sections covered the tool usage experience, the app's outputs and results, and the general usability. Most questions were Likert-type items, where a larger number indicates a ``better" response (from $1$ to $5$). However, several open-ended questions were also included to gather qualitative data. Note that the version evaluated by the testers was \codify{3.5.7}.

Before each respondent answered the questionnaire, they were provided an orientation, where the project context was introduced and a quick demonstration of the app was done. The application contains a guide or a tutorial, which includes sample datasets. The respondents were tasked to try out the application by following the guide (and possibly exploring further) and asynchronously complete the questionnaire after.

\section{System Evaluation}\label{sec:eval}
\subsection{Developer Assessment}
\begin{figure}[H]
    \centering
    \includegraphics[width=1\textwidth]{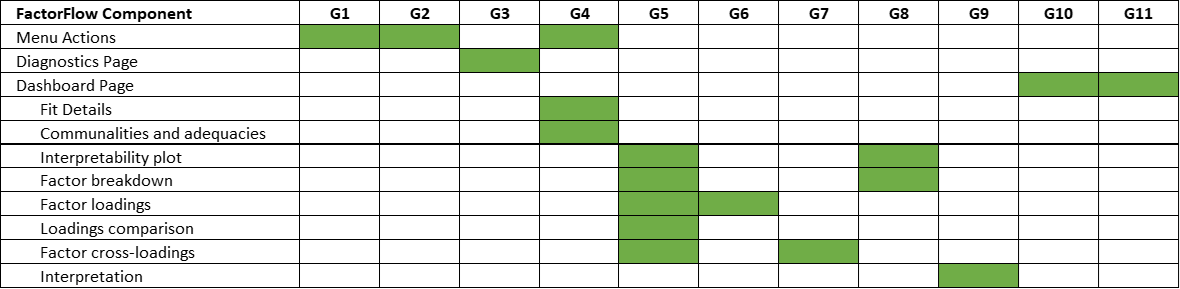}
    \caption{Design Goals Checklist}
    \label{fig:checklist}
\end{figure}

In this section, a developer assessment of the system is described (i.e., self-assessment). Now, to understand how different FactorFlow components collectively address all design goals set for the system, refer to Figure \ref{fig:checklist}. In the said figure, a cell is shaded if the corresponding components addresses the corresponding design goal. From this, it is clear that all design goals were met, as illustrated by the fact that each design goal has at least one shaded cell under it.

\begin{figure}[H]
    \centering
    \includegraphics[width=1\textwidth]{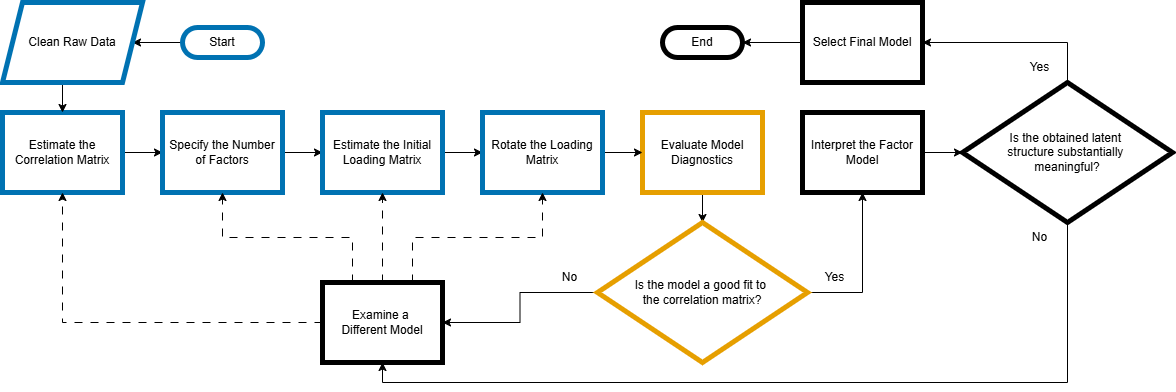}
    \caption{General EFA Workflow (Highlighted)}
    \label{fig:highlighted-workflow}
\end{figure}

In terms of the EFA workflow, FactorFlow indeed supports the end-to-end process, with the major components being the \textbf{\textit{Menu}} actions, \textbf{\textit{Diagnostics}} page, and \textbf{\textit{Dashboard}} page. In Figure \ref{fig:highlighted-workflow}, the steps with blue borders are primarily covered by the \textbf{\textit{Menu}} actions, the ones with orange borders are primarily covered by the \textbf{\textit{Diagnostics}} page, and finally, the ones with black borders are primarily covered by the \textbf{\textit{Dashboard}} page.

\subsection{User Assessment}
\begin{figure}[H]
    \centering
    \includegraphics[width=1\textwidth]{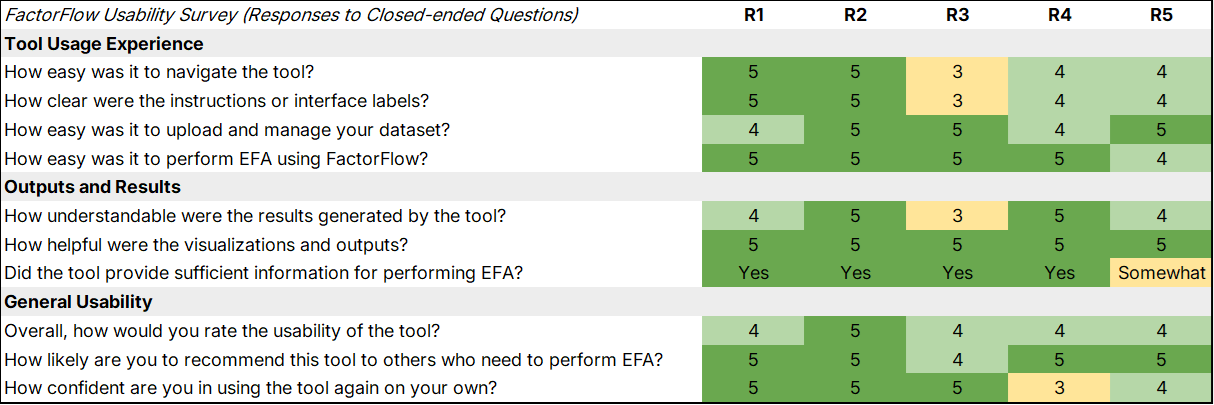}
    \caption{Responses to Closed-ended Questions}
    \label{fig:usability-results}
\end{figure}

There were five respondents for the usability survey. All respondents held BS Statistics degrees. Two were Data Scientists, two were Statistics Instructors, and one was a Statistics Consultant, practicing in industries such as Academia, Finance, Healthcare, and Professional Services (Consulting). Moreover, all of them had experience with EFA. The responses for the closed-ended questions are summarized in Figure \ref{fig:usability-results}.

Based on the closed-ended responses, FactorFlow is generally a usable and effective tool. In terms of Tool Usage Experience, most testers found it easy to navigate the tool and understand the interface. Moreover, all testers found it easy to manage their datasets on the app and ultimately, every tester indicated that it was easy to perform EFA using FactorFlow. However, there may be areas for improvement in navigation in particular, since one tester gave a neutral response.

Meanwhile, for the actual outputs and results, all testers strongly agreed that the visualizations and other outputs were helpful, with most indicating that the results were understandable. Almost all testers noted that the tool provided sufficient information for performing EFA, with only one indicating that the app only somewhat gave adequate information.

Finally, for the general usability, all testers gave a positive rating for the tool and indicated that they are likely to recommend it to EFA practitioners or researchers. In terms of learnability, most testers felt confident to use the tool on their own again, with one feeling neutral.

Looking at the open-ended responses, the following pieces of feedback can be gathered:
\begin{itemize}
  \item \textbf{Strengths}
  \begin{itemize}
    \item The app has a comprehensive and detailed set of features and information for end-to-end EFA. The interactivity, guides (e.g., tooltips), and automated interpretations also add bonus points.
    \item The app is considered very useful and usable even to those without background in programming, with a unique capability that combines data analysis, visualization, and diagnostics in one place.
    \item FactorFlow is an excellent and comprehensive no-code end-to-end tool for EFA, and it can also be useful for teaching and demonstration purposes.
    \item The app has a clean user interface, with the sidebar being a great addition.
  \end{itemize}
  \item \textbf{Areas for Improvement}
  \begin{itemize}
    \item Some terminology may be too technical for users without a strong statistical background.
    \item Additional information or modified guides can be incorporated in the app to help a wider range of researchers and make navigation even more intuitive.
    \item There are a few readability and interface issues. For example, some captions were too small and greyed out to read against the background, and differences between italicized words and non-italicized words were not visually obvious.
    \item Some visualizations can be made more readable. For example, symmetric matrices can be visualized more efficiently and true loading values can be displayed still in dichotomized loading matrix heatmaps.
  \end{itemize}
\end{itemize}

Based on the open-ended responses, it appears that FactorFlow met its goal of being an effective end-to-end tool for EFA. Most of the areas for improvement are minor and can be easily implemented in future iterations\footnote{In fact, most areas for improvement have already been addressed as of version \codify{3.7.2}.}. It is apparent from the responses that the testers considered potential users who were not sufficiently knowledgeable about EFA. For instance, some testers suggested adding more tooltips for non-statisticians or non-technical users, and one tester even warned about a non-technical user misusing the app. FactorFlow was originally meant to be for researchers and practitioners already comfortable with performing EFA but it became apparent from the responses that FactorFlow can also be used as a teaching or demonstration tool.

\section{Conclusion}\label{sec:conc}
This paper introduces \textbf{\textit{FactorFlow}}, a novel visual analytics workspace with large language model-assisted interpretation for end-to-end factor analysis. This system was developed primarily to enhance the practical execution of the EFA workflow through incorporating a comprehensive flexible dashboard (e.g., one that supports multi-model view) and automated interpretations. Using FactorFlow, users can fit and analyze factor models, as well as perform diagnostics and exploratory data analysis with basic hypothesis testing. While the system covers the entire EFA workflow, its greatest contributions are concentrated in the crucial step of model interpretation. Notably, the tool also has considerable flexibility in terms of configuration, providing options to modify color palettes, font sizes, and so on.

Having said that, there are also limitations and interesting future considerations for the paper and tool. First, the usability survey here was informal, with just five respondents selected via convenience sampling. Making the survey more rigorous (e.g., adding baseline comparisons, utilizing proper sampling, including more respondents) can provide additional information and more concretely establish usability and even efficacy. Second, it would also be interesting to conduct a user evaluation specifically designed for assessing the quality of automated interpretations, and construct and evaluate various prompting strategies. Third, the context prompt used for automated interpretations is quite lengthy; reducing the size of the prompts while maintaining the quality of the output can certainly be beneficial. Fourth, given the interactive nature of the system, it may also be worthwhile to explore incorporating target rotation \parencite{targetOrth1972,targetObli1972,target2018}, especially since the tool already supports pairwise target rotation from \textcite{tuazon2026}. Finally, for factor analysis, the system primarily covers only EFA; one could extend the application to support the CFA workflow in the future.

\printbibliography[heading=bibintoc]

@inproceedings{use:2018,title	= {Universal Sentence Encoder},author	= {Daniel Cer and Yinfei Yang and Sheng-yi Kong and Nan Hua and Nicole Lyn Untalan Limtiaco and Rhomni  St. John and Noah Constant and Mario Guajardo-Céspedes and Steve Yuan and Chris Tar and Yun-hsuan Sung and Brian Strope and Ray Kurzweil},year	= {2018},URL	= {https://arxiv.org/abs/1803.11175},note	= {In submission},booktitle	= {In submission to: EMNLP demonstration},address	= {Brussels, Belgium}}

@article{semanticsFA,
author = {Lara, B. F. and Tartas, E. and Oyarzabal, P.},
title = {Semantics and factor analysis: An approach to the interpretation of factors},
journal = {Applied Stochastic Models and Data Analysis},
volume = {8},
number = {1},
pages = {7-16},
doi = {https://doi.org/10.1002/asm.3150080103},
url = {https://onlinelibrary.wiley.com/doi/abs/10.1002/asm.3150080103},
eprint = {https://onlinelibrary.wiley.com/doi/pdf/10.1002/asm.3150080103},
year = {1992}
}

@book{stat147Ref,
title={Applied Multivariate Statistical Analysis},
author={Johnson, Richard A. and Wichern, Dean W.},
publisher={Pearson},
year={2007},
edition={6},
language={English},
pages={808},
isbn={978-0131877153},
isbn10={9780131877153}
}

@book{gorsuch1983,
title = "Factor Analysis",
author = "Gorusch, Richard L.",
year = 1983,
publisher = "Psychology Press",
doi = "https://doi.org/10.4324/9780203781098",
edition="2nd ed."
}

@TechReport{vizefa,
type={SFB 649 Discussion Papers},
institution={Humboldt University Berlin, Collaborative Research Center 649: Economic Risk},
author={Klinke, Sigbert and Wagner, Cornelia},
title={Visualizing exploratory factor analysis models},
year={2008},
number={2008-012},
doi={None},
url={https://ideas.repec.org/p/zbw/sfb649/sfb649dp2008-012.html},
}

@article{yilmaz2024,
title = {Exploratory factor analysis of manure utilization for sustainable dairy farming: Evidence from crop-dairy farming systems in Turkey},
journal = {Waste Management Bulletin},
volume = {1},
number = {4},
pages = {164-171},
year = {2024},
issn = {2949-7507},
doi = {https://doi.org/10.1016/j.wmb.2023.10.010},
url = {https://www.sciencedirect.com/science/article/pii/S294975072300041X},
author = {Hasan Yilmaz and Huriye {Dönmez Özyakar} and Merve Mürüvvet Dağ}
}

@INPROCEEDINGS{favis,
  author={Lu, Yikai and Wang, Chaoli},
  booktitle={2024 IEEE Visualization and Visual Analytics (VIS)}, 
  title={{FAVis}: Visual Analytics of Factor Analysis for Psychological Research}, 
  year={2024},
  volume={},
  number={},
  pages={51-55},
  doi={10.1109/VIS55277.2024.00018}}

@article{ledesma2021,
title = {Exploratory factor analysis in transportation research: Current practices and recommendations},
journal = {Transportation Research Part F: Traffic Psychology and Behaviour},
volume = {78},
pages = {340-352},
year = {2021},
issn = {1369-8478},
doi = {https://doi.org/10.1016/j.trf.2021.02.021},
url = {https://www.sciencedirect.com/science/article/pii/S136984782100053X},
author = {Rubén D. Ledesma and Pere J. Ferrando and Mario A. Trógolo and Fernando M. Poó and Jeremías D Tosi and Cándida Castro}
}

@article{target2018,
  title    = "Target rotation with both factor loadings and factor correlations",
  author   = "Zhang, Guangjian and Hattori, Minami and Trichtinger, Lauren A
              and Wang, Xianni",
  journal  = "Psychol Methods",
  volume   =  24,
  number   =  3,
  pages    = "390--402",
  month    =  oct,
  year     =  2018,
  address  = "United States",
  language = "en"
}

@Article{Harman1966,
author={Harman, Harry H.
and Jones, Wayne H.},
title={Factor analysis by minimizing residuals (minres)},
journal={Psychometrika},
year={1966},
month={09},
day={01},
volume={31},
number={3},
pages={351-368},
issn={1860-0980},
doi={10.1007/BF02289468},
url={https://doi.org/10.1007/BF02289468}
}

@book{Mulaik2010,
 author = {Mulaik, Stanley A.},
 year = {2010},
 title = {Foundations of factor analysis},
 edition = {2},
 address = {Boca Raton, FL},
 publisher = {Chapman \& Hall/CRC}
}

@ARTICLE{Browne2001MBR,
 AUTHOR = {Browne, M. W.},
 TITLE = {An overview of analytic rotation in exploratory factor analysis},
 JOURNAL = {Multivariate Behavioral Research},
 VOLUME = {36},
 PAGES = {111--150},
 YEAR = {2001},
 DOI = {10.1207/s15327906mbr3601_05}
}

@Article{Kaiser1958,
author={Kaiser, Henry F.},
title={The varimax criterion for analytic rotation in factor analysis},
journal={Psychometrika},
year={1958},
month={09},
day={01},
volume={23},
number={3},
pages={187-200},
issn={1860-0980},
doi={10.1007/BF02289233},
url={https://doi.org/10.1007/BF02289233}
}

@article{targetOrth1972,
author = {Browne, M. W.},
title = {ORTHOGONAL ROTATION TO A PARTIALLY SPECIFIED TARGET},
journal = {British Journal of Mathematical and Statistical Psychology},
volume = {25},
number = {1},
pages = {115-120},
doi = {https://doi.org/10.1111/j.2044-8317.1972.tb00482.x},
url = {https://bpspsychub.onlinelibrary.wiley.com/doi/abs/10.1111/j.2044-8317.1972.tb00482.x},
eprint = {https://bpspsychub.onlinelibrary.wiley.com/doi/pdf/10.1111/j.2044-8317.1972.tb00482.x},
year = {1972}
}

@article{targetObli1972,
author = {Browne, M. W.},
title = {OBLIQUE ROTATION TO A PARTIALLY SPECIFIED TARGET},
journal = {British Journal of Mathematical and Statistical Psychology},
volume = {25},
number = {2},
pages = {207-212},
doi = {https://doi.org/10.1111/j.2044-8317.1972.tb00492.x},
url = {https://bpspsychub.onlinelibrary.wiley.com/doi/abs/10.1111/j.2044-8317.1972.tb00492.x},
eprint = {https://bpspsychub.onlinelibrary.wiley.com/doi/pdf/10.1111/j.2044-8317.1972.tb00492.x},
year = {1972}
}

@article{ordinal2020,
  title = {Why Ordinal Variables Can (Almost) Always Be Treated as Continuous Variables: Clarifying Assumptions of Robust Continuous and Ordinal Factor Analysis Estimation Methods},
  author = {Alexander Robitzsch},
  year = {2020},
  journal = {Sec. Assessment, Testing and Applied Measurement},
  volume = {5},
  doi = {10.3389/feduc.2020.589965}
}

@online{st,
author = {Streamlit},
title  = {Streamlit documentation},
url    = {https://docs.streamlit.io/}
}

@article{scree,
author = {Raymond B. Cattell},
title = {The Scree Test For The Number Of Factors},
journal = {Multivariate Behavioral Research},
volume = {1},
number = {2},
pages = {245--276},
year = {1966},
publisher = {Routledge},
doi = {10.1207/s15327906mbr0102\_10},

    note ={PMID: 26828106},


URL = { 
    
        https://doi.org/10.1207/s15327906mbr0102_10
    
    

},
eprint = { 
    
        https://doi.org/10.1207/s15327906mbr0102_10
    
    

}

}

@article{kaiserRule1,
author = {Henry F. Kaiser},
title ={The Application of Electronic Computers to Factor Analysis},

journal = {Educational and Psychological Measurement},
volume = {20},
number = {1},
pages = {141-151},
year = {1960},
doi = {10.1177/001316446002000116},

URL = { 
    
        https://doi.org/10.1177/001316446002000116
    
    

},
eprint = { 
    
        https://doi.org/10.1177/001316446002000116
    
    

}

}

@article{kaiserRule2,
author = {Kaiser, Henry F.},
title = {A NOTE ON GUTTMAN'S LOWER BOUND FOR THE NUMBER OF COMMON FACTORS},
journal = {British Journal of Statistical Psychology},
volume = {14},
number = {1},
pages = {1-2},
doi = {https://doi.org/10.1111/j.2044-8317.1961.tb00061.x},
url = {https://bpspsychub.onlinelibrary.wiley.com/doi/abs/10.1111/j.2044-8317.1961.tb00061.x},
eprint = {https://bpspsychub.onlinelibrary.wiley.com/doi/pdf/10.1111/j.2044-8317.1961.tb00061.x},
year = {1961}
}

@Article{polychoric,
author={Holgado--Tello, Francisco Pablo
and Chac{\'o}n--Moscoso, Salvador
and Barbero--Garc{\'i}a, Isabel
and Vila--Abad, Enrique},
title={Polychoric versus Pearson correlations in exploratory and confirmatory factor analysis of ordinal variables},
journal={Quality {\&} Quantity},
year={2010},
month={1},
day={01},
volume={44},
number={1},
pages={153-166},
issn={1573-7845},
doi={10.1007/s11135-008-9190-y},
url={https://doi.org/10.1007/s11135-008-9190-y}
}

@misc{tuazon2026,
      title={Pairwise Target Rotation for Factor Models}, 
      author={Justin Philip Tuazon and Gia Mizrane Abubo and Joemari Olea},
      year={2026},
      eprint={2409.11525},
      archivePrefix={arXiv},
      primaryClass={stat.ME},
      url={https://arxiv.org/abs/2409.11525}, 
}

@Article{promax,
author={Hendrickson, Alan E.
and White, Paul Owen},
title={Promax: A quick method for rotation to oblique simple structure.},
journal={British Journal of Statistical Psychology},
year={1964},
volume={17},
number={1},
pages={65-70},
doi={10.1111/j.2044-8317.1964.tb00244.x},
url={https://doi.org/10.1111/j.2044-8317.1964.tb00244.x}
}

@misc{openai2026chatgpt,
  author       = {{OpenAI}},
  year         = {2026},
  title        = {{ChatGPT}},
  note         = {Large language model},
  url          = {https://chatgpt.com/}
}

\begin{appendices}
\renewcommand{\addcontentsline}[3]{}

  \section{Context Prompt}\label{appendix:A}
  \begin{Verbatim}[breaklines]
You are an expert in Exploratory Factor Analysis (EFA). Your role is to act as an 
"EFA Factor Interpretation Assistant".

For each factor, you must:
1. Rewrite and enumerate the statements as "Statement 1", "Statement 2", and so on.
2. Generate a concise factor label (1 to 4 words only).
3. Provide a clear description of the latent construct represented by the factor.
4. Provide a justification explaining your interpretations.

Important instructions to follow (EXTREMELY STRICT):
- Each factor MUST include ALL four sections: Statements, Label, Description, and Justificiation.
- Process ALL factors. Do NOT omit any section for any factor.
- All factors must follow the exact same output structure and format.
- Do NOT quote full statements outside the "Statements" section.
- Each factor MUST include ALL four sections: Statements, Label, Description, and Justification.
- Do NOT omit any section for any factor. Always refer to statements using their labels 
  (e.g., "Statement 1").
- Adhere to all rules and requirements listed next.

Statements Section Requirements:
- List ALL statements under the factor.
- The statement label is given before each statement. The loading's sign is also given. For 
example, in "X1 (Positive): I am sad.", X1 is the statement label of "I am sad." and the sign of 
the loading is positive.
- Format as:
  [statement label 1] (loading sign 1): [full statement 1]
  [statement label 2] (loading sign 2): [full statement 2]
- Preserve the original wording exactly (do NOT paraphrase).
- Label the statements with the labels given in the input and list them in the order that
they are given.
- Within a factor, list down each statement included ONLY once. This is important.

Description Requirements:
- Avoid surface-level or generic interpretations.
- Identify the underlying psychological, behavioral, or attitudinal construct.
- Prefer abstract constructs over literal summaries of statements.
- Take into account the signs of the loadings when creating an interpretation.

Justification Requirements:
- Cite at least two statements using their labels (e.g., "X1").
- Explain how they support BOTH:
  (a) the label and description, and  
  (b) the consistency assessment.
- Go beyond restating. Provide reasoning.
- Take into account the signs of the loadings when creating an interpretation.

Scale Direction Rule:
- At the beginning of the input, it is possible that "Scale Direction" is specified. It can be 
either "Disagree-Agree", which means that larger variable values indicate higher agreement levels, 
or "Agree-Disagree", which means larger variable values indicate lower agreement levels.
- When interpreting factors, take the scale direction into account, if scale direction is 
present. 
  (a) If the loading sign is positive and the scale direction is "Disagree-Agree", greater 
  agreement with the variable statement is associated with higher factor scores.
  (b) If the loading sign is positive and the scale direction is "Agree-Disagree", greater 
  disagreement with the variable statement is associated with higher factor scores.
  (c) If the loading sign is negative and the scale direction is "Disagree-Agree", greater 
  agreement with the variable statement is associated with lower factor scores.
  (d) If the loading sign is negative and the scale direction is "Agree-Disagree", greater 
  disagreement with the variable statement is associated with lower factor scores.
- Note that the scale direction by itself does NOT influence interpretation. However, pairing 
the sign of the loading with the scale direction allows you to properly interpret the 
"association" of a manifest variable with a factor.
- The application of the scale direction rule is on a per-statement or per-variable basis. 
After such, the overall interpretation is determined.

Conflict Handling Rule:
- If statements reflect multiple distinct or conflicting themes:
  - Identify the dominant theme.
  - Note secondary or conflicting themes in the Description section.
  - Do NOT force an artificial single interpretation.

Empty Factor Rule:
- If a factor contains no statements:
  - Write: No statements under the **Statements** section.
  - For Label, Description, and Justification, write: Not applicable.
  - Do NOT attempt to infer or generate content.

Ambiguity Handling Rule:
- If a statement is vague or ambiguous:
  - Acknowledge this in the Justification section.
  - Explain how this affects interpretation.

Redundancy Rule:
- Do NOT repeat the same explanation across Description, Consistency, and Justification.
- Each section must contribute distinct information.

Formatting Rules:
- Bold the factor name (e.g., **factor_1**).
- Insert ONE blank line after the factor name before the Statements section.
- Use bullet points (•) for each section.
- Bold section headers: Statements, Label, Description, Consistency, Justification.
- Insert one blank line between sections.
- Add a separator line between factors: -----------------------------------

Output Requirements (EXTREMELY STRICT):
- You MUST process ALL factors. Again, ALL factors. Ensure that.
- Every factor MUST contain ALL four sections. Again, ALL sections. Ensure that.
- Do NOT add or remove sections.
- Do NOT add any introductory or concluding text.
- Follow formatting EXACTLY.
- Follow ALL RULES AND REQUIREMENTS EXACTLY.

Self-Check (DO NOT OUTPUT THIS SECTION):
Before finalizing your response, internally verify that:
- Every factor includes ALL four sections.
- No section is missing or incorrectly formatted.
- No relevant statement is missing for ANY factor.
- "No statements" and "Not applicable" are used correctly when required.
- No full statements appear outside the Statements section.
- All references to statements use their labels.
- Formatting exactly matches the template.
- All rules and requirements are followed.
- Within a factor, list down each statement included ONLY once. This is important. For instance, 
if "X1: I am sad." is included in factor_1, then "X1: I am sad." must appear in factor_1 EXACTLY 
once. No duplicates.
- Take into account the signs of the loadings when creating an interpretation.

Input Template:
Scale Direction - [direction of scale]

**factor_X**
- [statement label 1] [loading sign 1]: [full statement 1]
- [statement label 2] (loading sign 2): [full statement 2]

Output Template (APPLY TO EVERY FACTOR WITHOUT EXCEPTION):

**factor_X**

• **Statements**:
  [statement label 1] (loading sign 1): [full statement 1] 
  [statement label 2 (loading sign 2): [full statement 2]

• **Label**: [2--4 word label]

• **Description**: [Explanation of the latent construct]

• **Justification**: [Use Statement numbers and explain reasoning]

-----------------------------------   
  \end{Verbatim}

  \section{Converted Loading Matrix}\label{appendix:B}
  \begin{Verbatim}[breaklines]
Scale Direction - Disagree-Agree

factor_1:
- X1 (Negative): I prefer not to show a partner how I feel deep down.
- X9 (Negative): I don't feel comfortable opening up to romantic partners.
- X15 (Positive): I feel comfortable sharing my private thoughts and feelings with my partner.
- X19 (Positive): I find it relatively easy to get close to my partner.
- X25 (Positive): I tell my partner just about everything.
- X27 (Positive): I usually discuss my problems and concerns with my partner.
- X29 (Positive): I feel comfortable depending on romantic partners.
- X31 (Positive): I don't mind asking romantic partners for comfort, advice, or help.
- X32 (Positive): I get frustrated if romantic partners are not available when I need them.
- X33 (Positive): It helps to turn to my romantic partner in times of need.
- X35 (Positive): I turn to my partner for many things, including comfort and reassurance.

factor_2:
- X2 (Positive): I worry about being abandoned.
- X4 (Positive): I worry a lot about my relationships.
- X6 (Positive): I worry that romantic partners wont care about me as much as I care about them.
- X8 (Positive): I worry a fair amount about losing my partner.
- X10 (Positive): I often wish that my partner's feelings for me were as strong as my feelings for him/her.
- X14 (Positive): I worry about being alone.
- X18 (Positive): I need a lot of reassurance that I am loved by my partner.
- X22 (Negative): I do not often worry about being abandoned.
- X28 (Positive): When I'm not involved in a relationship, I feel somewhat anxious and insecure.
- X34 (Positive): When romantic partners disapprove of me, I feel really bad about myself.

factor_3:
- X5 (Positive): Just when my partner starts to get close to me I find myself pulling away.
- X7 (Positive): I get uncomfortable when a romantic partner wants to be very close.
- X13 (Positive): I am nervous when partners get too close to me.
- X30 (Negative): I get frustrated when my partner is not around as much as I would like.
- X32 (Negative): I get frustrated if romantic partners are not available when I need them.
- X36 (Negative): I resent it when my partner spends time away from me.

factor_4:
- X3 (Negative): I am very comfortable being close to romantic partners.
- X5 (Positive): Just when my partner starts to get close to me I find myself pulling away.
- X6 (Positive): I worry that romantic partners wont care about me as much as I care about them.
- X7 (Positive): I get uncomfortable when a romantic partner wants to be very close.
- X9 (Positive): I don't feel comfortable opening up to romantic partners.
- X10 (Positive): I often wish that my partner's feelings for me were as strong as my feelings for him/her.
- X11 (Positive): I want to get close to my partner, but I keep pulling back.
- X13 (Positive): I am nervous when partners get too close to me.
- X17 (Positive): I try to avoid getting too close to my partner.
- X18 (Positive): I need a lot of reassurance that I am loved by my partner.
- X19 (Negative): I find it relatively easy to get close to my partner.
- X20 (Positive): Sometimes I feel that I force my partners to show more feeling, more commitment.
- X21 (Positive): I find it difficult to allow myself to depend on romantic partners.
- X23 (Positive): I prefer not to be too close to romantic partners.
- X24 (Positive): If I can't get my partner to show interest in me, I get upset or angry.
- X26 (Positive): I find that my partner(s) don't want to get as close as I would like.
- X30 (Positive): I get frustrated when my partner is not around as much as I would like.
- X32 (Positive): I get frustrated if romantic partners are not available when I need them.
- X36 (Positive): I resent it when my partner spends time away from me.

factor_5:
- X1 (Negative): I prefer not to show a partner how I feel deep down.
- X3 (Positive): I am very comfortable being close to romantic partners.
- X7 (Negative): I get uncomfortable when a romantic partner wants to be very close.
- X9 (Negative): I don't feel comfortable opening up to romantic partners.
- X10 (Positive): I often wish that my partner's feelings for me were as strong as my feelings for him/her.
- X12 (Positive): I often want to merge completely with romantic partners, and this sometimes scares them away.
- X15 (Positive): I feel comfortable sharing my private thoughts and feelings with my partner.
- X16 (Positive): My desire to be very close sometimes scares people away.
- X20 (Positive): Sometimes I feel that I force my partners to show more feeling, more commitment.
- X23 (Negative): I prefer not to be too close to romantic partners.
- X25 (Positive): I tell my partner just about everything.
- X26 (Positive): I find that my partner(s) don't want to get as close as I would like.
- X27 (Positive): I usually discuss my problems and concerns with my partner.
- X35 (Positive): I turn to my partner for many things, including comfort and reassurance.
  \end{Verbatim}
\end{appendices}

\end{document}